\documentclass[prd,superscriptaddress,unsortedaddress,twocolumn,showpacs,floatfix]{revtex4}

\usepackage{epsfig}
\usepackage{dcolumn}
\usepackage{amsmath}

\newcommand{\KLpienu}{\mbox{$K_{L}\to\pi^{\pm}e^{\mp}\nu$}}
\newcommand{\KLpilnu}{\mbox{$K_{L}\to\pi^{\pm}\ell^{\mp}\nu$}}
\newcommand{\KLpimunu}{\mbox{$K_{L}\to\pi^{\pm}{\mu}^{\mp}\nu$}}
\newcommand{\pzgg}{\mbox{$\pi^0\rightarrow \gamma\gamma$}}
\newcommand{\pz}{\mbox{$\pi^0$}}

\newcommand{\Rzz}{\mbox{$\Gamma(K_{L}\rightarrow \pi^0\pi^0)$}}
\newcommand{\Rzzz}{\mbox{$\Gamma(K_{L}\rightarrow \pi^0\pi^0\pi^0)$}}

\newcommand{\Rpm}{\mbox{$\Gamma(K_{L}\rightarrow \pi^+\pi^-)$}}
\newcommand{\Rpienu}{\mbox{$\Gamma(\KLpienu)$}}
\newcommand{\Rpimunu}{\mbox{$\Gamma(\KLpimunu)$}}
\newcommand{\Rppp}{\mbox{$\Gamma(K_{L}\rightarrow \pi^+\pi^-\pi^0)$}}

\newcommand{\Gzz}{\mbox{$\Gamma_{00}$}}
\newcommand{\Gzzz}{\mbox{$\Gamma_{000}$}}
\newcommand{\Gpm}{\mbox{$\Gamma_{+-}$}}
\newcommand{\Gpienu}{\mbox{$\Gamma_{Ke3}$}}
\newcommand{\Gpimunu}{\mbox{$\Gamma_{K\mu3}$}}
\newcommand{\Gppp}{\mbox{$\Gamma_{+-0}$}}
 \newcommand{\beq}{\begin{equation}}
 \newcommand{\eeq}{\end{equation}}
 \newcommand{\bqa}{\begin{eqnarray}}
 \newcommand{\eqa}{\end{eqnarray}}

\newcommand{\K}{\mbox{$K$}}

\newcommand{\KL}{\mbox{$K_{L}$}}
\newcommand{\KS}{\mbox{$K_{S}$}}

\newcommand{\brems}{Bremsstrahlung}

\def\epe{\epsilon'\!/\epsilon}

\newcommand{\epsrat}{\mbox{$\epsilon^{\prime}\!/\epsilon$}}
\newcommand{\reepoe}{\mbox{$Re(\epsrat)$}}

\newcommand{\ktev}{\mbox{KTeV}}
\newcommand{\etapm}{\mbox{$\eta_{+-}$}}

\def\kpi0{K_{L}\to 3\pi^0}
\def\ke3{K_{L}\to\pi^{\pm}e^{\mp}\nu}
\def\k2pi{K_{L} \to \pi^+\pi^-}

\newcommand{\KLpp}{\mbox{$K_{L}\rightarrow\pi\pi$}}

\newcommand{\Kethree}{\mbox{$K_{e3}$}}
\newcommand{\Kmuthree}{\mbox{$K_{\mu 3}$}}
\newcommand{\Kthreepi}{\mbox{$K_{+-0}$}}
\newcommand{\Ktwopi}{\mbox{$K_{+-}$}}

\newcommand{\KLpm}{\mbox{$K_{L}\rightarrow\pi^{+}\pi^{-}$}}

\newcommand{\KLzz}{\mbox{$K_{L}\rightarrow\pi^{0}\pi^{0}$}}

\newcommand{\KLzzz}{\mbox{$K_{L}\rightarrow \pi^{0}\pi^{0}\pi^{0}$}}
\newcommand{\KLpmz}{\mbox{$K_{L}\rightarrow \pi^{+}\pi^{-}\pi^{0}$}}

\newcommand{\Lppi}{\mbox{$\Lambda \rightarrow p \pi^-$}}

\newcommand{\eeg}{\mbox{$e^+e^-\gamma$}}

\newcommand{\ppc}{\mbox{$\pi^{+}\pi^{-}$}}

\newcommand{\ppn}{\mbox{$\pi^{0}\pi^{0}$}}

\newcommand{\pmz}{\mbox{$\pi^{+}\pi^{-}\pi^{0}$}}
\newcommand{\zzz}{\mbox{$\pi^{0}\pi^{0}\pi^{0}$}}
\newcommand{\zz}{\mbox{$\pi^{0}\pi^{0}$}}

\newcommand{\ring}{{\tt RING}}

\newcommand{\eu}{ \times 10^{-4}}
\newcommand{\uptsq}{ {\rm GeV}^2/c^2 }
\newcommand{\umass}{ {\rm GeV}/c^2 }

\newcommand\vacveto{Vacuum-Veto}
\newcommand\specveto{Spec-Veto}
\newcommand\csiveto{CsI-Veto}

\newcommand{\ppzkin}{\mbox{$k_{+-0} $}}
\newcommand{\magvus}{\mbox{$\vert V_{us}\vert$}}
\newcommand{\mageta}{\mbox{$\vert \eta_{+-}\vert$}}

\newcommand{\ptsq}{\mbox{$p_t^2$}}

\newcommand{\mpp}{\mbox{$m_{\pi\pi}$}}
\newcommand{\sqmpp}{\mbox{$m^2_{\pi\pi}$}}
\newcommand{\EK}{\mbox{$E_K$}}
\newcommand{\ZK}{\mbox{$Z_K$}}
\newcommand{\pchi}{\mbox{$\chi^2_{\pi^0}$}}
\newcommand{\dray}{\mbox{$\delta$-ray}}

\newcommand{\drbrem}{\mbox{$\Delta R_{\gamma {\rm brem}}$}}

\newcommand{\BLpm}{\mbox{$B_{\pi^+\pi^-}^L$}}
\newcommand{\BLzz}{\mbox{$B_{\pi^0\pi^0}^L$}}
\newcommand{\BSklthree}{\mbox{$B_{\pi\ell\nu}^S$}}

\newcommand{\TLvalue}{\mbox{$5.15$}}
\newcommand{\TLerror}{\mbox{$0.04$}}

\def\RPMNvalue{0.6640}
\def\RPMNerrstat{0.0014}
\def\RPMNerrsyst{0.0022}
\def\RPMNerrtot{0.0026}

\def\NdataPMNa{   394300}
\def\NdataPMNb{   449379}

\def\accPMNa{ 0.239}
\def\accPMNb{ 0.180}

\def\RZZZvalue{0.4782}
\def\RZZZerrstat{0.0014}
\def\RZZZerrsyst{0.0053}
\def\RZZZerrtot{0.0055}

\def\NdataZZZa{   209365}
\def\NdataZZZb{   211950}

\def\accZZZa{ 0.046}
\def\accZZZb{ 0.022}

\def\RPPPvalue{0.3078}
\def\RPPPerrstat{0.0005}
\def\RPPPerrsyst{0.0017}
\def\RPPPerrtot{0.0018}

\def\NdataPPPa{   799501}
\def\NdataPPPb{   807343}

\def\accPPPa{ 0.200}
\def\accPPPb{ 0.124}

\def\RPPvalue{ 4.856}
\def\RPPerrstat{ 0.017}
\def\RPPerrsyst{ 0.023}
\def\RPPerrtot{ 0.029}

\def\NdataPPa{    83725}
\def\NdataPPb{   979799}

\def\accPPa{ 0.265}
\def\accPPb{ 0.121}

\def\RNEUTvalue{ 4.446}
\def\RNEUTerrstat{ 0.016}
\def\RNEUTerrsyst{ 0.019}
\def\RNEUTerrtot{ 0.025}

\def\NdataNEUTa{   100365}
\def\NdataNEUTb{  1609324}

\def\accNEUTa{ 0.150}
\def\accNEUTb{ 0.054}

\def\lovshich{   3.4}

\def\BKEvalue{0.4067}

\def\BKEerrtot{0.0011}
\def\WKEvalue{0.7897}

\def\WKEerrtot{0.0065}
  
\def\BKMvalue{0.2701}

\def\BKMerrtot{0.0009}
\def\WKMvalue{0.5244}

\def\WKMerrtot{0.0044}
  
\def\BZZZvalue{0.1945}

\def\BZZZerrtot{0.0018}
\def\WZZZvalue{0.3777}

\def\WZZZerrtot{0.0045}
  
\def\BPMZvalue{0.1252}

\def\BPMZerrtot{0.0007}
\def\WPMZvalue{0.2431}

\def\WPMZerrtot{0.0023}
  
\def\BPMvalue{ 1.975}

\def\BPMerrtot{ 0.012}
\def\WPMvalue{ 3.835}

\def\WPMerrtot{ 0.038}
  
\def\BZZvalue{ 0.865}

\def\BZZerrtot{ 0.010}
\def\WZZvalue{ 1.679}

\def\WZZerrtot{ 0.024}
  
\def\etapmv  { 2.228}
\def\etapme  { 0.010}
\def\etapmint{ 0.005}
\def\etapmext{ 0.009}
\def\kspmzz  { 2.261}

\def\kspmzze { 0.033}

\def\ikrbv{0.6622}

\def\LeptUni{0.9969}
\def\LeptUniE{0.0048}
\def\DeltaRat{1.0058}
\def\DeltaRatE{0.0010}

\def\ikrbetot{0.0018}

\begin{document}

\title{
       Measurements of $K_L$ Branching Fractions
       and the CP Violation Parameter $|\eta_{+-}|$
    }
\newcommand{\UAz}{University of Arizona, Tucson, Arizona 85721}
\newcommand{\UCLA}{University of California at Los Angeles, Los Angeles,
                    California 90095} 
\newcommand{\UCSD}{University of California at San Diego, La Jolla,
                   California 92093} 
\newcommand{\EFI}{The Enrico Fermi Institute, The University of Chicago, 
                  Chicago, Illinois 60637}
\newcommand{\UB}{University of Colorado, Boulder, Colorado 80309}
\newcommand{\ELM}{Elmhurst College, Elmhurst, Illinois 60126}
\newcommand{\FNAL}{Fermi National Accelerator Laboratory, 
                   Batavia, Illinois 60510}
\newcommand{\Osaka}{Osaka University, Toyonaka, Osaka 560-0043 Japan} 
\newcommand{\Rice}{Rice University, Houston, Texas 77005}
\newcommand{\UVa}{The Department of Physics and Institute of Nuclear and 
                  Particle Physics, University of Virginia, 
                  Charlottesville, Virginia 22901}
\newcommand{\UW}{University of Wisconsin, Madison, Wisconsin 53706}

\affiliation{\UAz}
\affiliation{\UCLA}
\affiliation{\UCSD}
\affiliation{\EFI}
\affiliation{\UB}
\affiliation{\ELM}
\affiliation{\FNAL}
\affiliation{\Osaka}
\affiliation{\Rice}
\affiliation{\UVa}
\affiliation{\UW}

\author{T.~Alexopoulos}   \affiliation{\UW}
\author{M.~Arenton}       \affiliation{\UVa}
\author{R.F.~Barbosa}     \altaffiliation[Permanent address: ]
   {University of S\~{a}o Paulo, S\~{a}o Paulo, Brazil}\affiliation{\FNAL}
\author{A.R.~Barker}      \altaffiliation[Deceased.]{ } \affiliation{\UB}
\author{L.~Bellantoni}    \affiliation{\FNAL}
\author{A.~Bellavance}    \affiliation{\Rice}
\author{E.~Blucher}       \affiliation{\EFI}
\author{G.J.~Bock}        \affiliation{\FNAL}
\author{E.~Cheu}          \affiliation{\UAz}
\author{S.~Childress}     \affiliation{\FNAL}
\author{R.~Coleman}       \affiliation{\FNAL}
\author{M.D.~Corcoran}    \affiliation{\Rice}
\author{B.~Cox}           \affiliation{\UVa}
\author{A.R.~Erwin}       \affiliation{\UW}
\author{R.~Ford}          \affiliation{\FNAL}
\author{A.~Glazov}        \affiliation{\EFI}
\author{A.~Golossanov}    \affiliation{\UVa}
\author{J.~Graham}        \affiliation{\EFI}   
\author{J.~Hamm}          \affiliation{\UAz}
 
\author{K.~Hanagaki}      \affiliation{\Osaka}
\author{Y.B.~Hsiung}      \affiliation{\FNAL}
\author{H.~Huang}         \affiliation{\UB}
\author{V.~Jejer}         \affiliation{\UVa}  
\author{D.A.~Jensen}      \affiliation{\FNAL}
\author{R.~Kessler}       \affiliation{\EFI}
\author{H.G.E.~Kobrak}    \affiliation{\UCSD}
\author{K.~Kotera}        \affiliation{\Osaka}
\author{J.~LaDue}         \affiliation{\UB}
  
\author{A.~Ledovskoy}     \affiliation{\UVa}
\author{P.L.~McBride}     \affiliation{\FNAL}

\author{E.~Monnier}
   \altaffiliation[Permanent address: ]{C.P.P. 
    Marseille/C.N.R.S., France}\affiliation{\EFI}
\author{H.~Nguyen}       \affiliation{\FNAL}
\author{R.~Niclasen}     \affiliation{\UB} 
\author{V.~Prasad}       \affiliation{\EFI}
\author{X.R.~Qi}         \affiliation{\FNAL}
\author{E.J.~Ramberg}    \affiliation{\FNAL}
\author{R.E.~Ray}        \affiliation{\FNAL}
\author{M.~Ronquest}	 \affiliation{\UVa}
\author{E. Santos}       \altaffiliation[Permanent address: ]
      {University of S\~{a}o Paulo, S\~{a}o Paulo, Brazil}\affiliation{\FNAL}
\author{P.~Shanahan}     \affiliation{\FNAL}
\author{J.~Shields}      \affiliation{\UVa}
\author{W.~Slater}       \affiliation{\UCLA}
\author{D.~Smith}	 \affiliation{\UVa}
\author{N.~Solomey}      \affiliation{\EFI}
\author{E.C.~Swallow}    \affiliation{\EFI}\affiliation{\ELM}
\author{P.A.~Toale}      \affiliation{\UB}
\author{R.~Tschirhart}   \affiliation{\FNAL}
\author{Y.W.~Wah}        \affiliation{\EFI}
\author{J.~Wang}         \affiliation{\UAz}
\author{H.B.~White}      \affiliation{\FNAL}
\author{J.~Whitmore}     \affiliation{\FNAL}
\author{M.~Wilking}      \affiliation{\UB}
\author{B.~Winstein}     \affiliation{\EFI}
\author{R.~Winston}      \affiliation{\EFI}
\author{E.T.~Worcester}  \affiliation{\EFI}
\author{T.~Yamanaka}     \affiliation{\Osaka}
\author{E.~D.~Zimmerman} \affiliation{\UB}

\collaboration{The KTeV Collaboration}


\vspace*{1.cm}

\begin{abstract}
  We present new measurements of the six largest branching fractions of the
  $K_L$ using data collected in 1997 by the 
  {\ktev} experiment (E832) at Fermilab. The results are
$B(\KLpienu)= \BKEvalue \pm \BKEerrtot$, 
$B(\KLpimunu)= \BKMvalue \pm \BKMerrtot$,
$B(\KLpmz)= \BPMZvalue \pm \BPMZerrtot $,
$B(\KLzzz)= \BZZZvalue \pm \BZZZerrtot $,
$B(\KLpm)=(\BPMvalue \pm \BPMerrtot)\times 10^{-3}$, and
$B(\KLzz)=(\BZZvalue \pm \BZZerrtot)\times 10^{-3} $,
  where statistical and systematic errors have been summed in quadrature. 
 We also determine the CP violation parameter $\mageta$ to be
 $(\etapmv \pm \etapme)\times 10^{-3}$.  Several of these results are not
in good agreement with averages of previous measurements.
\end{abstract}


\maketitle





  \section{Introduction}
  \label{sec:intro}


Most recent experimental work on the $\KL$ has focused on 
rare and CP violating decays.  The branching fractions of the main
$\KL$ decay modes, however, have not been measured together
in a modern, high-statistics experiment.
These branching fractions are
fundamental experimental parameters used to determine the CKM element
$\magvus$, the CP violation parameter $\mageta$, and the normalization
for many other rare decay measurements.

In this paper, we present new results for the six largest 
$K_L$ branching fractions:
\KLpienu, \KLpimunu, \KLpmz, \KLzzz, \KLpm, and \KLzz. 
We determine these branching fractions
by measuring the following ratios of decay rates:
\begin{eqnarray}
\Gpimunu/\Gpienu & \equiv & \Rpimunu/ \Rpienu  \\
\Gppp/\Gpienu    & \equiv & \Rppp/ \Rpienu\  \ \ \\ 
\Gzzz/\Gpienu    & \equiv & \Rzzz/ \Rpienu  \\
\Gpm/\Gpienu     & \equiv & \Rpm/ \Rpienu  \\
\Gzz/\Gzzz       & \equiv & \Rzz/ \Rzzz. 
\end{eqnarray}
Each ratio is measured in a statistically independent data sample 
collected by the \ktev\ (E832) experiment at Fermilab.
Note that throughout this paper, inner bremsstrahlung contributions
are included for all decay modes with charged particles.

Since the six decay modes listed above account for more than
99.9\% of the total decay rate, the five partial width ratios 
may be converted into branching fraction measurements.
For example, the \KLpienu\ branching fraction, $B_{Ke3}$, 
may be written as
\begin{equation}
  B_{Ke3}=\frac{1-B_{rare}} 
     {  1
        +\frac{\Gpimunu}{\Gpienu}
        +\frac{\Gzzz}{\Gpienu}
        +\frac{\Gppp}{\Gpienu}
        +\frac{\Gpm}{\Gpienu}
        +\frac{\Gzz}{\Gpienu}
    }, 
\label{eq:ke3}
\end{equation}
where $B_{rare}=0.07\%$ is the sum of branching fractions of
other rare $\KL$ decay modes~\footnote{The
     contribution of rare decays to the $\KL$ total width is mostly
     from $B(\KL \to \gamma\gamma) = 0.060\%$, 
     $B(\KL \to \pi^0\pi^{\pm} e^{\mp} \nu)=0.005\%$,
     and the direct emission $B(\KL \to \pi^+\pi^-\gamma)= 0.002\%$
     \cite{pdg02}. The invisible width is assumed to be zero.}.
In terms of our measured partial width ratios,
$\Gzz/\Gpienu = \Gzz/\Gzzz \times \Gzzz/\Gpienu$.

The paper is organized as follows. 
First, we give a brief description of 
the KTeV detector and the data sets used in this analysis. 
Next, we present an overview of the analysis techniques followed 
by a more detailed discussion of selection criteria for the 
individual decay modes.
Section~\ref{sec:MC} contains a description of the Monte Carlo simulation
used to determine the detector acceptance, and a discussion of the resulting
systematic uncertainties. 
In Section~\ref{sec:results}, we present results for the 
partial width ratios and branching fractions,
along with several crosschecks of our analysis. Our 
branching fraction measurements 
are then used to extract $\mageta$.  
Finally, our results are compared with previous measurements.


   \section{\ktev\ detector}
   \label{sec:detector}

\begin{figure*}
\centering
\psfig{figure=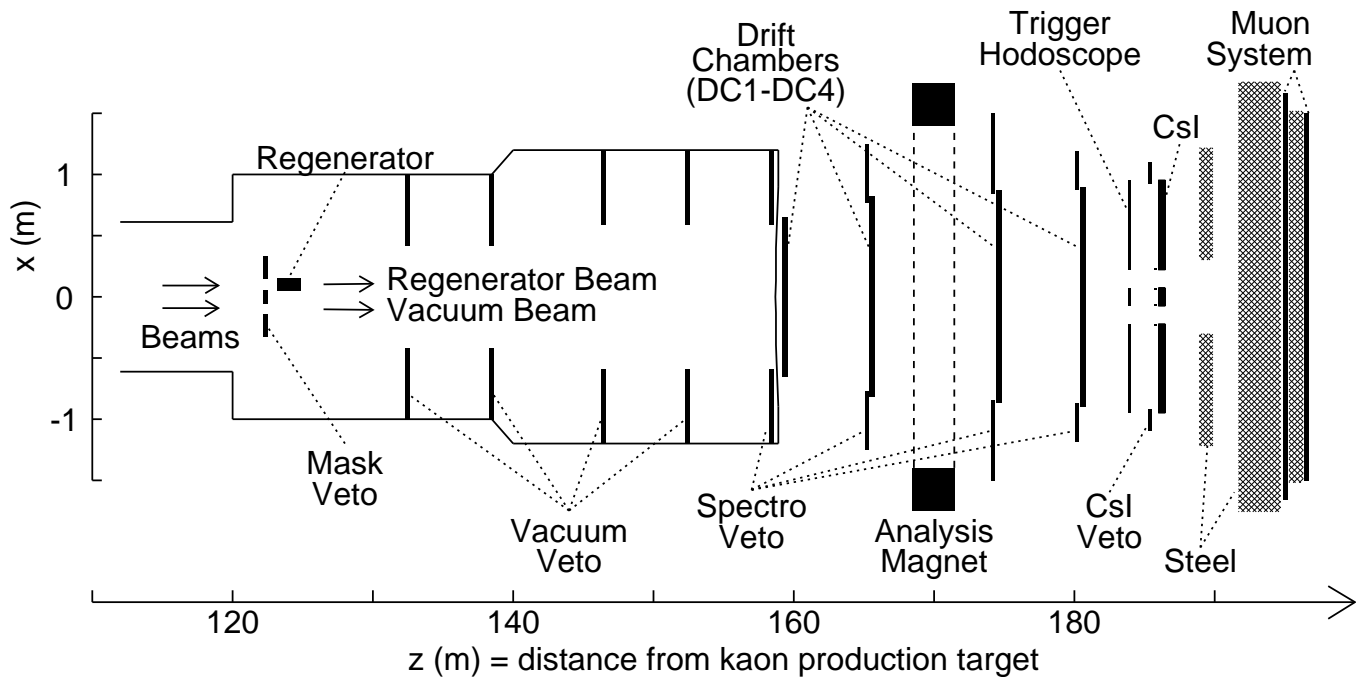,width=\linewidth}
\caption{
    Plan view of the KTeV (E832) detector.
    The evacuated decay volume ends with a thin vacuum window at
    $Z = 159$~m.  
    The label ``CsI'' indicates the electromagnetic
    calorimeter.
        }
\label{fig:detector}
\end{figure*}

The \ktev\ detector (see Fig.~\ref{fig:detector})
and associated event reconstruction techniques have 
been described in detail elsewhere~\cite{Alavi-Harati:2002ye}. 
Here we give a brief summary of the essential detector components.
An 800 GeV/c proton beam striking a BeO target is used to produce two
almost parallel neutral beams, shaped by a series of collimators. 
Except for a special ``low-intensity'' run described later,
a fully-active regenerator is placed in 
one of the beams 
to provide a source of $K_S$ for
the measurement of $\epe$; decays in the ``regenerator'' beam are 
not used in this analysis. The other beam, referred to as the vacuum beam,
provides $K_L$ decays used for these measurements.  
A large vacuum decay region 
extends to 159 m from the primary target.

Following a thin vacuum window at the end of the vacuum 
region is a drift chamber spectrometer  used to measure
the momentum of charged particles;
this spectrometer consists of four chambers, 
two upstream (DC1-DC2) and 
two downstream (DC3-DC4) of an analysis magnet that imparts a 0.41 GeV/$c$ 
momentum kick in the horizontal plane.
Each chamber measures positions in both the $x$ and $y$ 
views (transverse to the beam direction).
Farther downstream lies a trigger (scintillator) hodoscope 
to identify charged particles,
and a 3100 crystal,
pure cesium iodide (CsI) electromagnetic calorimeter.
Downstream of the CsI calorimeter there is a muon system consisting
of scintillator hodoscopes
behind 4 m and 5 m of steel.
Veto detectors surround
the vacuum decay region, each drift chamber, and the
CsI calorimeter (\vacveto, \specveto, 
and \csiveto).

KTeV uses a three-level trigger system to reduce the total rate 
of recorded events.  The Level 1 and Level 2 triggers are implemented in hardware and
the Level 3 trigger is a software filter that uses the full 
event reconstruction.
With the exception of \KLzz,
the data samples used in this analysis do not require the Level
3 trigger.

The analysis of the $K_L$ branching fractions 
benefits from the KTeV detector design that was optimized to measure
the direct CP violation parameter  
using  a Monte Carlo simulation (MC) to determine the
detector acceptance. 
To reduce uncertainties in the simulation, it is important that
apertures and detector geometry are well measured, 
and that there is very little material before the calorimeter 
to affect decay products.

The $Z$-locations of detector elements are known with
100 $\mu$m precision. 
The transverse sizes
of the drift chambers are known to $20$~$\mu$m; the
transverse dimensions and relative locations of the other detector elements
are determined to better than $200$~$\mu$m using charged particle tracks. 

The detector has very little material upstream of the calorimeter, reducing
losses from multiple scattering, hadronic interactions, and $\gamma \to
e^+e^-$ conversions.
The material from the vacuum window to the last drift chamber
is only 0.012 radiation lengths ($X_0$),
or 0.007 pion interaction length ($\Lambda_0$);
the material downstream of the last drift chamber to the front face of the
CsI is $0.031~X_0$, or  $0.014~\Lambda_0$.

The analysis presented in this paper also benefits from extensive detector
calibration performed for the $\epe$ analysis~\cite{Alavi-Harati:2002ye}.
The momentum kick of the analysis magnet
is determined to 0.01\%\ precision using $\KLpm$ events 
and the PDG value of the kaon mass~\cite{pdg02}.
The CsI calorimeter energy scale is determined to better than 0.1\% 
based on calibration using 500 million momentum analyzed electrons 
from \KLpienu\ events.
The momentum resolution of the spectrometer and the 
electromagnetic energy resolution of the CsI calorimeter 
are both better than 1\%.

  \section{Data Collection}
  \label{sec:data}


In this paper, we report results based on data taken during two periods
in 1997, which will be referred to as ``high intensity'' and 
``low intensity.''
All five partial width ratios are measured in both the high and low 
intensity
data samples. The quoted result for each ratio is based on the sample 
yielding the smaller total 
uncertainty; the other sample is used as a crosscheck.
With the exception of $\Gpimunu/\Gpienu$, the high intensity sample
gives a smaller uncertainty
\footnote{$\Gpimunu/\Gpienu$ has poor statistics in the high intensity 
      sample because the only trigger which does not veto muons was
      pre-scaled by 10,000. }.

The high intensity period was used primarily to collect $\K\to \pi\pi$ 
decays for the \reepoe\ measurement \cite{Alavi-Harati:2002ye}.
In addition to the nominal 2-pion triggers, several additional triggers
with relaxed requirements were included for systematic studies;
these additional triggers provide the data samples for the 
branching fraction analysis\footnote{For the decay modes 
with branching fractions greater than 10\% (semileptonic and $3\pi$),
even the relaxed, highly prescaled triggers have high statistics.
To get sufficient statistics in the $K\to\pi\pi$ modes (with $BR \sim 0.1$\%),
we use the same 2-pion triggers as in the  $\epe$ analysis.
For \Gpm/\Gpienu, we use the standard Level~1 and Level~2 trigger,
but use the pre-scaled ($1/100$) sample in which there is no Level~3.
For \Gzz/\Gzzz, we use 5\% of the nominal sample in the $\epe$ analysis.}.

The low intensity data were collected during a 2-day special run.
The beam intensity was lowered by a factor of 10,
and the drift chamber operating voltage was raised to increase efficiency.  
In addition, the regenerator was removed, resulting in two vacuum
beams and eliminating extra 
detector hits from the interaction of beam particles
in the regenerator.
The data collected during the low intensity period have significantly
lower detector activity.  For example,
the average number of spurious
drift chamber hits is only $2.3$ in the low intensity period 
compared to $43$ for the high intensity period.


   \section{Measurement Strategy}
   \label{sec:measure}


The analysis is optimized to reduce systematic uncertainties 
resulting from the Monte Carlo simulation
used to correct for acceptance differences between pairs of decay modes.
With the exception of $\Gzzz/\Gpienu$, 
we consider ratios of decay modes  with
similar final state particles.
To be insensitive to the absolute trigger efficiency,
events for the numerator and denominator of each partial width ratio
are collected with a single trigger (the only exception is the 
measurement of $\Gzz/\Gzzz$
\footnote{The measurement of \Gzz/\Gzzz\ is the only partial 
          width ratio based on data from two different triggers. 
          This analysis uses the same four and six cluster 
          triggers used in the $\epe$ analysis.
          For the low intensity sample, used as a crosscheck,  
          both decay modes are collected with  the same neutral-mode 
          minimum bias trigger, but the low number of \KLzz\ events 
          in this sample
          limits the precision of the measurement.
     }).

The trigger requirements for each partial width ratio are summarized in 
Table~\ref{tb:trigger}. 
The main trigger requirement for each pair of
decay modes is either two charged tracks or a large energy deposit
in the CsI calorimeter. For $\Gzzz /\Gpienu$, the trigger requirement is 
only 25 GeV of energy in the calorimeter for both decay modes; there is 
no charged-track requirement for $\KLpienu$.

\begin{table}[hb]
\caption{
     Trigger requirements used to measure each
     partial width ratio. Note that two different triggers are used to 
     measure \Gzz/\Gzzz. ``Two charged tracks'' refers to hits in the
     drift chambers and trigger hodoscope, ``total CsI energy'' requires
     more than 25 GeV energy-sum in the 3100 channels, 
     ``CsI clusters'' refers to the number of clusters above 1~GeV,
     and ``vetos'' are used to reject events. 
        }
   \label{tb:trigger}
\begin{ruledtabular}
\begin{tabular}{lcccccc}
    partial          & two      & total     &          &  Vac,  &       &      \\
    width            & charged  & CsI       & CsI      &  Spec  & CsI   &  muon \\ 
    ratio            & tracks   & energy    & clusters &  vetos & veto  &  veto \\
\hline 
 \Gpimunu/\Gpienu    &   yes    &  no       &  --      &  yes   & no    & no  \\
 \Gppp/\Gpienu       &   yes    &  no       &  --      &  no    & no    & yes \\
 \Gpm/\Gpienu        &   yes    &  no       &  --      &  yes   & yes   & yes \\
 \Gzzz/\Gpienu       &   no     &  yes      &  --      &  yes   & no    & yes \\
\Gzz\ for \Gzz/\Gzzz &   no     &  yes      &   4      &  yes   & yes   & yes \\
\Gzzz\ for \Gzz/\Gzzz&   no     &  yes      &   6      &  yes   & yes   & yes \\
\hline 
\end{tabular}
\end{ruledtabular}
\end{table}

As will be described below, very simple event selection requirements may be
used to distinguish different kaon decay modes from each other, and to reduce 
background to a negligible level for all decay modes.  For some decay modes,
the excellent spectrometer and calorimeter resolution allows us to achieve
this background rejection without using all of the available detector
information. 
For the $\KLpimunu$ and $\KLpmz$ decay modes, we exploit this flexibility 
to reduce systematic uncertainties in the acceptance.

For the \KLpimunu\ decay mode, we do not make use of the muon system
to identify the muon; this avoids systematic errors in modeling muon 
propagation in the steel in front of the muon hodoscope as well as in 
modeling gaps between the muon counters.
For the \KLpmz\ decay mode, $\pzgg$ is not reconstructed in order to
be insensitive to pion showers that bias the 
photon energy measurement in the CsI calorimeter.
Also, by ignoring the \pz, the reconstruction of 
\KLpmz\ and \KLpienu\ decays
are very similar, reducing the uncertainty on the 
acceptance ratio of these two modes.
As a crosscheck (Section~\ref{subsec:crosscheck}), 
the $\KLpimunu$ and $\KLpmz$ modes are also analyzed
using the muon system and fully reconstructing 
the $\pi^0$.

All reconstructed decay modes are required to have kaon energy, 
$\EK$, between 40 and 120 GeV, and decay position, $\ZK$,  
between 123 and 158 m from the target.
For the reconstruction of semileptonic and \KLpmz\ decays, 
there is a missing particle ($\nu$ or $\pz$);
this leads to multiple kaon energy solutions.
All energy solutions are required to be in the accepted range. 
These $\EK$ and $\ZK$ ranges are more restrictive than in the
$\epe$ analysis in order to 
have a negligible contribution from
$\KS \to \pi\pi$ decays.

In the following two sections, we discuss the techniques used to 
reconstruct ``charged'' decay modes with two oppositely-charged particles,
and ``neutral'' decay modes with only photons in the final state.
More details of the KTeV event reconstruction are given 
in~\cite{Alavi-Harati:2002ye}.


   \section{Charged Decay Mode Analysis}
   \label{sec:charged}


\subsection{Charged Decay Mode Reconstruction and Event Selection}

The reconstruction of \KLpienu, \KLpimunu, \KLpmz, and \KLpm\ 
begins with the identification of two oppositely charged tracks
coming from a single vertex. 
To pass the event selection, one of the two tracks must 
be within $7$~cm of a CsI cluster; the second
track is not required to have a cluster match. As will
be discussed later, this relaxed track-cluster matching requirement
reduces the inefficiency arising from hadronic
interactions upstream of the calorimeter.

The transverse ($X,Y$) decay vertex is required to be within an 
11$\times$11 cm$^2$ square centered on the beam
($\ring < 121~{\rm cm}^2$~\footnote{\ring\ is defined as the area,
in cm$^2$, of the smallest square centered on the beam that contains
the transverse decay vertex at the $Z$ position of the CsI calorimeter.
For all-neutral events, $\ring$ is defined in terms of the 
center-of-energy measured in the calorimeter~\cite{Alavi-Harati:2002ye}.
               }). 
The beam profile is about 10$\times$10 cm$^2$ at the CsI calorimeter.
This cut removes most events in which the kaon has scattered
in a collimator.

The fiducial region for the charged decay modes
is defined by requiring that projections of tracks fall safely within 
the boundaries of the drift chambers, trigger hodoscope, 
and the CsI calorimeter.
The tracks are also subject to a
``cell separation'' cut~\cite{Alavi-Harati:2002ye}, which requires
that the tracks never share the same drift chamber cell. This requirement 
introduces an effective inner aperture, rejecting
pairs of tracks with a very small opening angle.
At the CsI calorimeter, the tracks are
required to have a large 40 cm separation
to minimize the overlap of hadronic and electromagnetic showers.

\begin{figure}[hb]
 \epsfig{file=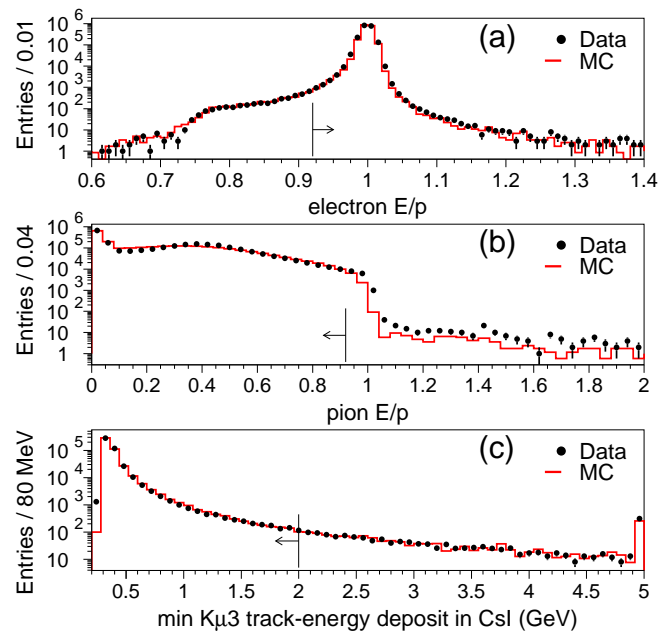, width=\linewidth}
\caption{
    (a) $E/p$ (calorimeter energy over spectrometer momentum)
         distribution for electrons; 
    (b) $E/p$ distribution for pions; 
    (c)  minimum cluster energy deposit among two tracks for 
         \Kmuthree\ decays (events with energy deposits
         greater than 5~GeV are shown in the last bin).
         The arrows indicate selection requirements.
       }
    \label{fig:pid} 
\end{figure}

The different charged decay modes are distinguished from each other
on the basis of particle identification  and kinematics.
The calorimeter energy measurement ($E$), 
combined with the spectrometer momentum ($p$),
is used to distinguish electrons and pions.
Figures~\ref{fig:pid}(a) and (b) 
show data and Monte Carlo (MC) $E/p$ distributions for
electrons and pions, respectively.
Electron candidates are required to have
$E/p$ greater than 0.92; this cut retains $99.8\%$ 
of the electrons and rejects $99.5\%$ of the pions.
Pions are required to have $E/p$ less than 0.92.

In the \KLpimunu\ analysis, 
the $E/p$ requirement is used
to reject \Kethree\ decays.
We also require that
at least one track point to a CsI cluster 
with energy less than 2~GeV
(Fig.~\ref{fig:pid}(c)).
The 2~GeV cluster requirement retains 99.7\% 
of the \Kmuthree\ signal. 
This requirement does not
distinguish the pion from the muon in 1/3 of the events,
because 1/3 of the pions do not shower in the CsI calorimeter.
Since the pion and muon are not identified,
there are four $E_K$ solutions that are all required to
be within the 40-120 GeV range.
Recall that muons are not identified with the muon system 
in order to reduce acceptance uncertainties;
for the other decay mode analyses, however,
the muon system is used in veto to suppress background
from \Kmuthree\ and pion decays.

In addition to particle identification for the two tracks,
kinematic requirements are also used to distinguish
the charged decay modes.
We use three variables ($\mpp$, $\ptsq$, and $\ppzkin$), 
each computed under the assumption
that both tracks are charged pions.
The only fully reconstructed decay, $\KLpm$, is isolated
using the two-track invariant mass ($\mpp$)
and using the two-track transverse momentum-squared ($\ptsq$)
measured with respect to the line connecting the primary target
and decay vertex. 
To separate \KLpmz\ from semileptonic decays, 
we use an additional kinematic variable,
\begin{equation}
\ppzkin = 
  \frac{\textstyle \left( m^2_K - {\sqmpp} - m^2_{\pi^0}\right)^2 - 
   4 {\sqmpp} m^2_{\pi^0} - 4 m^2_K \ptsq } 
   {\textstyle 4({\sqmpp} + \ptsq)  },
   \label{eq:pp0kin}
\end{equation}
where $m_K$ and $m_{\pi^0}$ are the kaon and $\pi^0$ masses, respectively.
For \KLpmz\ decays, \ppzkin\ corresponds to the square of the 
longitudinal momentum of the 
$\pi^0$ in the reference frame in which the sum of the charged pion momenta
is orthogonal to the kaon momentum.  

\begin{figure}
\epsfig{file=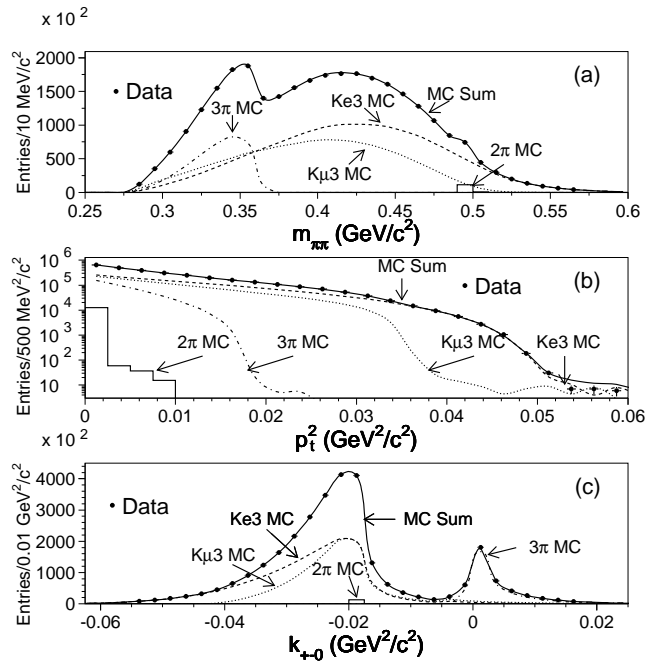, width=\linewidth}
\caption{
   For all two-track events, without particle identification or
   kinematic requirements, distributions are shown for
   (a) $\mpp$, (b) $\ptsq$, and (c) $\ppzkin$.
   Data are shown as dots. MC simulations for
   \KLpienu, 
   \KLpimunu,
   \KLpmz, and
   \KLpm\ are indicated.
   The sum of all MC contributions
   is shown as a solid line. 
       }
    \label{fig:rawkin} 
\end{figure}

To illustrate the use of these 3 variables, 
Fig.~\ref{fig:rawkin} shows data and MC
distributions of \mpp, \ptsq, and \ppzkin\ 
for all two-track events
before particle identification and kinematic requirements are applied.
The different MC samples are normalized to 
each other using the  branching fractions reported in this paper.
Figures~\ref{fig:rawkin}(a) and (b) show peaked distributions
in \mpp\ and \ptsq\ for \KLpm\ decays.
\KLpm\ candidates are selected with $0.488 < \mpp < 0.508~\umass$
and $\ptsq < 2.5\eu~\uptsq$; the combined efficiency
of these two requirements is 99.2\%.
For the other charged decay modes, the same \mpp\ requirement 
is used to veto background from \KLpm\ decays.

Figure~\ref{fig:rawkin}(c) shows the \ppzkin\ distribution.
There are two well-separated enhancements, corresponding to 
semileptonic and \KLpmz\ decays.
We require \KLpmz\ candidates to have
$\ppzkin > -0.005~\uptsq$, 
which retains $99.9\%$ of the \KLpmz\ decays.
For semileptonic decays, we require $\ppzkin < -0.006~\uptsq$; 
this keeps $99.5\%$ of \Kethree\ decays and 97.9\% of \Kmuthree\ decays.

In addition to misidentifying kaon decay modes, 
background can also arise from hyperon decays.
Background from $\Lppi$ decays is suppressed to a
negligible level for all charged decay modes by removing events 
with invariant $p\pi^-$ mass
consistent with $m_{\Lambda}$ ($1.112-1.120$~GeV/$c^2$);
to determine $m_{p\pi}$,
the higher momentum track is assumed to be the proton.

\begin{figure}
\epsfig{file=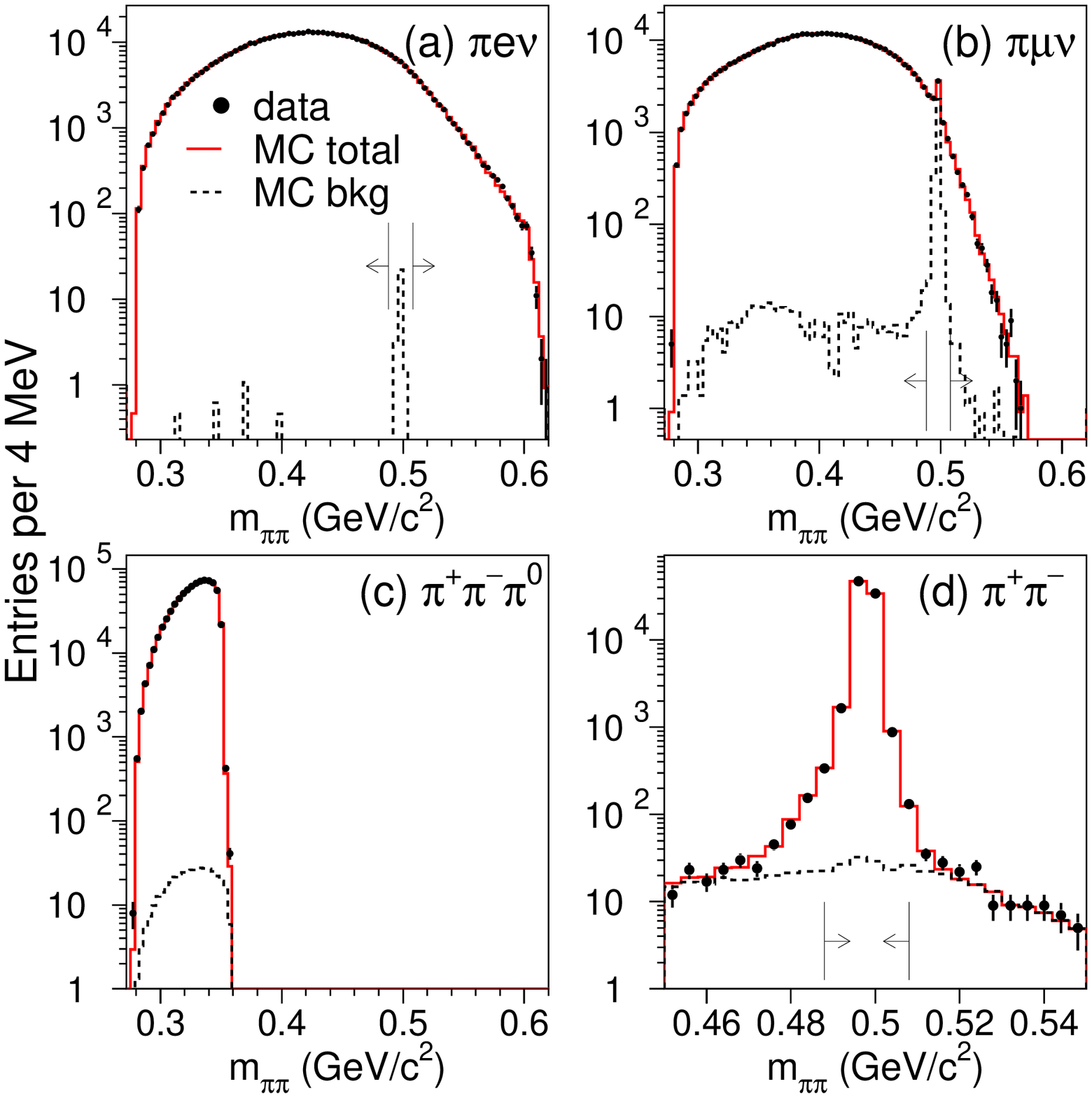, width=\linewidth}
\caption{
   $\mpp$ distributions
   after all analysis requirements except $\mpp$.
   Selection requirements are for
   (a) $\KLpienu$, (b) $\KLpimunu$
   (c) $\KLpmz$, (d) $\KLpm$.
   Data (MC) are shown by dots (histogram).
   ``MC total'' refers to signal plus background.
   ``MC bkg'' is the scattering+background prediction
   based on simulating the other three (non-signal)
   charged decay modes.
   The horizontal arrows indicate the region(s) selected by the
   \mpp\ requirement. 
   Note that the \mpp\ scale is different in (d).
       }
    \label{fig:m2pi} 
\end{figure}

\begin{figure}
\epsfig{file=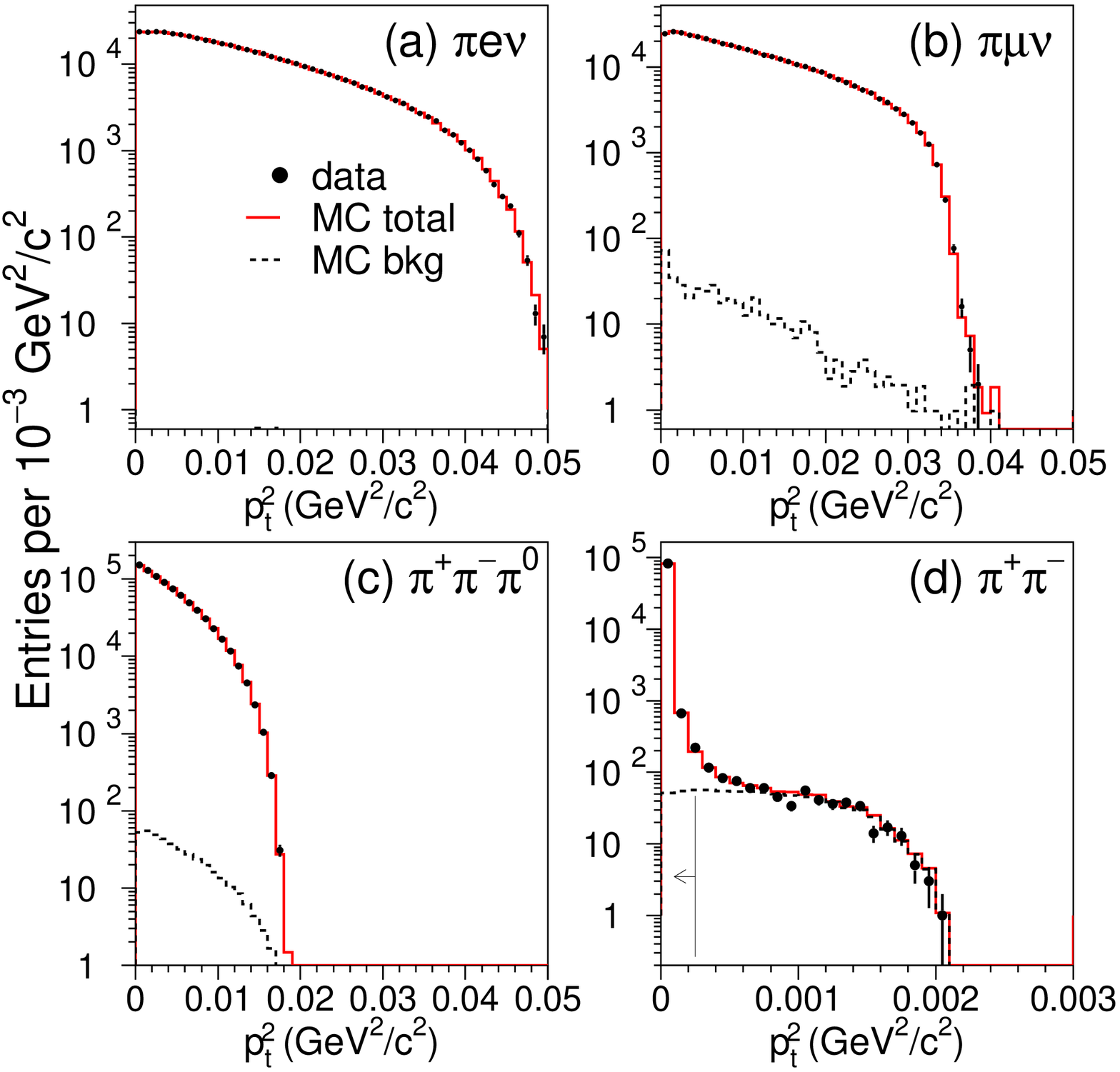, width=\linewidth}
\caption{
   $\ptsq$ distributions
   after all analysis requirements except $\ptsq$.
   Selection requirements are for
   (a) $\KLpienu$, (b) $\KLpimunu$
   (c) $\KLpmz$, (d) $\KLpm$.
   Data (MC) are shown by dots (histogram).
   ``MC total'' refers to signal plus background.
   ``MC bkg'' is the scattering+background prediction
   based on simulating the other three (non-signal)
   charged decay modes.
   The horizontal arrows indicate the region selected by the
   \ptsq\ requirement. 
   Note that the \ptsq\ scale is different in (d).
       }
    \label{fig:ptsq} 
\end{figure}

\begin{figure}[hb]
\epsfig{file=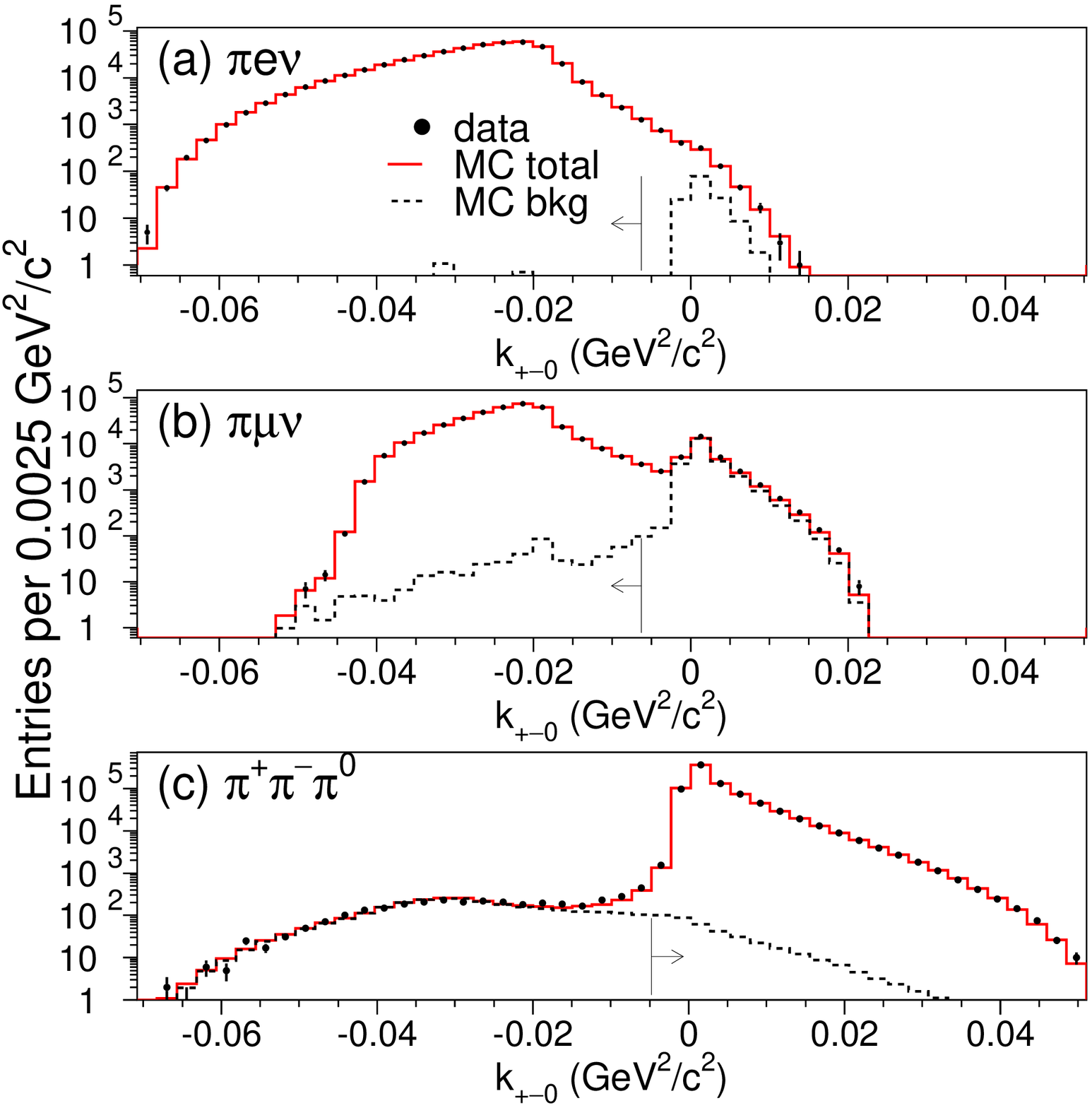, width=\linewidth}
\caption{
   $\ppzkin$ distributions
   after all analysis requirements except $\ppzkin$.
   Selection requirements are for
   (a) $\KLpienu$, (b) $\KLpimunu$, (c) $\KLpmz$.
   Data (MC) are shown by dots (histogram).
   ``MC total'' refers to signal plus background.
   ``MC bkg'' is the scattering+background prediction
   based on simulating the other three (non-signal)
   charged decay modes.
   The horizontal arrows indicate the region selected by the
   \ppzkin\ requirement. 
       }
    \label{fig:ppzkin} 
\end{figure}

Figures~\ref{fig:m2pi}-\ref{fig:ppzkin} show the 
data and MC distribution of
\mpp, \ptsq, and \ppzkin\ for each charged decay mode,
after applying all event selection requirements except for 
the requirement on the plotted variable.
The data and MC distributions agree well;
the background for each signal mode is based on 
simulating the other three  charged decay modes.

In addition to the general reconstruction and event selection 
described above, several notable 
features specific to certain decay modes are described below.

The \KLpienu\ decay mode is used in four of the five partial width ratios.
Some \KLpienu\ selection requirements are adjusted depending on 
the ratio.
For $\Gpimunu/\Gpienu$, the pion is required to satisfy a 
stricter $E/p$ requirement ($E/p < 0.85$ instead of 0.92) 
to reduce background from $\Kethree$ in the $\Kmuthree$ sample.
This stricter $E/p$ requirement retains 99.1\% of pions,
and rejects 99.93\% of the electrons.
Since there is one pion in the final state for both 
\Kmuthree\ and \Kethree,
we use the same strict $E/p$ requirement in both decay modes
to eliminate the systematic uncertainty from this 
requirement.

The measurement of $\Gzzz/\Gpienu$ is based on a trigger 
that requires $25$~GeV energy deposit in the CsI calorimeter. 
As discussed in Section~\ref{sec:measure},
the minimum $\EK$ requirement is 40~GeV, 
which is well above the 25~GeV trigger threshold.
The $\EK$ requirement ensures that the \KLzzz\ acceptance
is insensitive to the trigger threshold,
but is not sufficient 
for \Kethree\ decays because of the missing neutrino.
For the \Kethree\ acceptance to be insensitive to the trigger
threshold, we require that the electron momentum 
be above 34~GeV/$c$.

Although the $\KLpmz$ decay mode is selected without reconstructing
the \pz\ in the CsI calorimeter,
the \pz\ decay products can hit the veto detectors.
To eliminate the uncertainty in modeling the veto system efficiency,
$\Gppp/\Gpienu$ is measured using a trigger that does not
include the veto system.
Photon clusters from the \pz\ decay may also overlap a pion cluster
resulting in an $E/p$ measurement that is too high. 
To reduce the influence of this effect,
pion candidates are allowed to have either $E/p$ less than 0.92 or $E/p$
greater than 1.05. Allowing pions with $E/p > 1.05$ recovers 1.5\%\ of
the \KLpmz\ sample.

  \subsection{Charged Decay Mode Background}
  \label{subsec:chrg bkg}

Table~\ref{ta:chback} summarizes the background sources
 that are subtracted
from each charged decay mode.
The background to \Kethree\ decays depends on the specific selection
for the partial width ratio, but is always 
less than $3\times 10^{-5}$. 
The background in the other three charged decays is $\sim 10^{-3}$. 
For the partially reconstructed decay modes 
($\KLpienu$, $\KLpimunu$, and $\KLpmz$),
the background level is checked using the \ppzkin\ distributions
(Figure~\ref{fig:ppzkin});
based on the agreement of the data and Monte Carlo distributions,
we assign a 20\% systematic uncertainty
to the background levels.
The $\KLpm$ background is evaluated in the same manner as in 
the $\epe$ analysis
  \footnote{The background to \KLpm\ is 0.16\%, 
            which is larger than the 0.1\% background
            quoted in \cite{Alavi-Harati:2002ye}; 
            this is because the pion selection efficiency
            does not cancel in \Gpm/\Gpienu, and therefore
            the $E/p$ requirement is relaxed to reduce
            the systematic uncertainty.
         }.

There are also events in which the parent kaon has
scattered in the defining collimator.
This ``collimator scattering'' contribution includes a 
regenerated $K_S$-component,
and therefore must be subtracted in the \KLpm\ analysis;
collimator scattering is
suppressed to 0.01\% using the \ptsq\ requirement.
For the partially reconstructed decay modes,
collimator scattering is suppressed to 0.1\%\ 
using the \ring\ requirement;
this scattering component is not subtracted,
and is included in the samples for both data and MC.

\begin{table}
\caption{
    \label{ta:chback}
     Charged decay backgrounds from other kaon decays.
       }
\begin{ruledtabular}
\begin{tabular}{ l | cccc}
                 & \multicolumn{4}{c}{Background ($\times 10^4$) to:} \\
Decay Mode       & ~\Kethree\ ~& ~\Kmuthree\ ~& ~\Kthreepi\ ~& ~\Ktwopi\ ~ \\
\hline 
\Kethree\        & ---       & 3        &   2.4     &  10   \\
\Kmuthree\       & $0.02$    & ---      &   2.7     &   5   \\
\Kthreepi\       & $<0.1$    & 5        &  ---      &  ---    \\
\Ktwopi\         & $<0.2$    & 3        &  ---    
    &  1\footnote{This background is from collimator scattering.}    \\
\hline  
Total            & $<0.3$     & 11        &   5    &  16   \\ 
\end{tabular}
\end{ruledtabular}
\end{table}


 \section{Neutral Decay Mode Analysis}
 \label{sec:neutral}


  \subsection{Neutral Decay Mode Reconstruction and Event Selection}

The reconstruction of the $\KLzz$ and  $\KLzzz$  decay modes
is based on energies and positions of photons measured in the
CsI electromagnetic calorimeter as described in~\cite{Alavi-Harati:2002ye}. 
Exactly four (six) clusters,
each with  a transverse profile consistent with a photon,
are required for $\KLzz$ ($\KLzzz$).
The clusters must be separated from each other 
by at least $7.5$~cm and have energy greater than  $3$~GeV. 
The fiducial volume is defined by 
cluster positions measured in the calorimeter.  
We reject events in which any cluster position is reconstructed
in the layer of crystals adjacent to the beam holes 
(Fig.~\ref{fig:csihole}(a))
or in the outermost layer of crystals.

The center-of-energy of photon clusters is required
to lie within an 11$\times$11 cm$^2$ square centered on the 
beam profile ($\ring < 121~{\rm cm}^2$);
the \ring\ distribution for each neutral decay mode is 
shown in Fig.~\ref{fig:neutring}.
The \ring\ cut removes most events in which the kaon has scattered
in the collimator or regenerator.

Photons are paired to reconstruct two or three
neutral pions consistent with a single decay vertex. 
The number of possible photon pairings is 3 for
$\KLzz$ and 15 for $\KLzzz$. To select the best pairing,
we introduce a ``pairing-$\chi^2$'' variable ($\pchi$),  
which quantifies the consistency of the $\pi^0$ vertices.
The pairing $\chi^2$ is required to be less than 
50 for $\KLzz$  and less than 75 for $\KLzzz$
\footnote{
          In the $\epe$ analysis, the \pchi\ cuts are 12 (24)
          for $\KLzz$ ($\KLzzz$). 
          To measure partial width ratios with a different number of
          neutral pions in the numerator and denominator, 
          this \pchi\ requirement is relaxed
          to reduce acceptance uncertainties.
      }.  
This procedure identifies the correct photon pairing in
more than 99\%\ of the events.
The $\ZK$ location of the kaon decay vertex is determined from a 
weighted average of the  $\pz$ vertices. 
The main kinematic requirement is that the invariant mass of
the $2\pz$ or $3\pz$ final state (Fig.~\ref{fig:neutmass})
be between $0.490$ and $0.505~\umass$.

  \subsection{Neutral Decay Mode Backgrounds}
  \label{subsec:neut bkg}

The background subtraction procedure for $\KLzz$ is identical 
to that used in  the $\epe$ analysis.
The background composition is 
0.30\% from scattering in the regenerator,
0.09\% from collimator scattering,
and 0.32\% from \KLzzz\ in which two photons are undetected
(``$3\pz$-background'')
\footnote{
    The 0.32\% \zzz\ background in the \KLzz\ sample 
    is three times larger than in the $\epe$ analysis 
    \cite{Alavi-Harati:2002ye}
    because of the much looser \pchi\ cut.
     }.
The total background is $(0.71\pm 0.06)\%$.

\begin{figure}[hb]
\epsfig{file=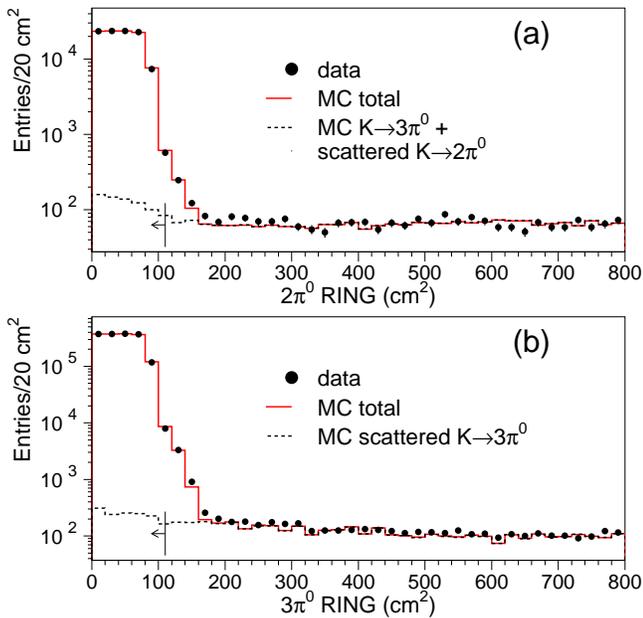, width=\linewidth}
\caption{
    \ring\ distribution for (a) \KLzz\ and (b) \KLzzz\ candidates.
    Data are shown with dots. ``MC total'' (histogram) refers
    to the simulation of the signal including 
    backgrounds and scattering in the collimator and regenerator.
    The dashed histogram shows scattering+background predicted by MC.
    The arrow indicates the analysis requirement.
       }
    \label{fig:neutring} 
\end{figure}

\begin{figure}[hb]
\epsfig{file=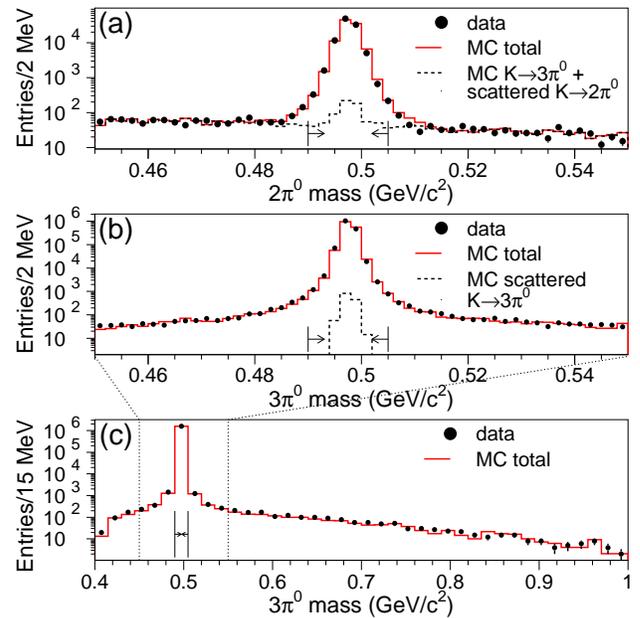, width=\linewidth}
\caption{
    (a) $2\pz$ mass, (b) $3\pz$ mass, and (c) $3\pz$ mass
    shown with extended mass scale.
    Data are shown with dots. ``MC total'' (histogram) refers
    to the simulation of the signal including 
    backgrounds scattering in the collimator and regenerator.
    The dashed histogram shows scattering+background predicted by MC.
    The arrows indicate the analysis requirement.
       }
    \label{fig:neutmass} 
\end{figure}

In Fig.~\ref{fig:neutring}(a), 
events with $\ring > 200~{\rm cm}^2$ are almost entirely
due to scattering in the regenerator and collimator; 
the MC predicts both the absolute level and \ring-shape.
Note that events with scattered kaons have the same invariant-mass
distribution as events with 
unscattered kaons, and therefore cannot be 
identified in
the \ppn\ mass distribution (Fig.~\ref{fig:neutmass}(a)).

In Fig.~\ref{fig:neutmass}(a), 97\%\ of the events outside the signal region
result from misreconstructed
$\KL \to 3\pz$ events and the remaining 3\%\ are from
from \KLzz\ events with the wrong photon pairing. These mass sidebands
are well modeled in the simulation.
The $3\pz$-background is responsible for the apparent
increase in background under the \ring-signal in 
Fig.~\ref{fig:neutring}(a). \\

In the \KLzzz\ decay mode, 
no source of background has been identified.
The contribution from kaon scattering (0.1\% after the \ring\ cut)
is not subtracted; 
it is well modeled in the simulation, as shown in 
Fig.~\ref{fig:neutring}(b).
The mass side-bands in Fig.~\ref{fig:neutmass}(b) and (c)
result from \KLzzz\ decays in which the wrong photon pairing is used 
to compute the invariant mass.
These mis-pairings, which are well modeled by the MC,
are not subtracted.

  \section{Monte Carlo simulation and Systematic Uncertainties}
  \label{sec:MC}


\begin{table*}[ht]
\caption{
    \label{tab:syst}
     Systematic uncertainties in partial width ratios (in percent)
       }
\begin{ruledtabular}
\begin{tabular}{lccccc}
\hline
  Source of uncertainty  & \Gpimunu/ \Gpienu  & \Gzzz/ \Gpienu  
& \Gppp/ \Gpienu & \Gpm/ \Gpienu & \Gzz/ \Gzzz  \\
\hline
 Acceptance (MC Simulation) & & & & & \\
 \ \ \ Event Generation:  & & & & & \\
\ \ \ - Kaon energy spectrum                                  &    0.02    &    0.16    &    0.04    &    0.02    &    0.01 \\
\ \ \ - Form factor                                           &    0.11    &    0.08    &    0.29    &    0.08    &    0.00 \\
 \ \ \ Radiative corrections:  & & & & & \\
\ \ \ -                                                       &    0.15    &    0.20    &    0.14    &    0.14    &    0.00 \\
 \ \ \ Particle Propagation:  & & & & & \\
\ \ \ - Detector material                                     &    0.10    &    0.56    &    0.33    &    0.33    &    0.15 \\
\ \ \ - Detector geometry                                     &    0.02    &    0.39    &    0.05    &    0.02    &    0.08 \\
 \ \ \ Detector Response:  & & & & & \\
\ \ \ - Accidental activity                                   &    0.00    &    0.22    &    0.04    &    0.02    &    0.03 \\
\ \ \ - Trigger                                               &    0.00    &    0.07    &    0.10    &    0.07    &    0.28 \\
\ \ \ - $e^{\pm}, \mu^{\pm}, \pi^{\pm}$ reconstruction        &    0.21    &    0.70    &    0.24    &    0.26    &    0.00 \\
\ \ \ - $\pi^0$ reconstruction                                &    0.00    &    0.37    &    0.00    &    0.00    &    0.23 \\
Background                                            &    0.10    &    0.00    &    0.02    &    0.04    &    0.04 \\
$B(\pi^0 \to \gamma\gamma)$                           &    0.00    &    0.10    &    0.10    &    0.00    &    0.03 \\
Monte Carlo Statistics                                &    0.10    &    0.12    &    0.05    &    0.13    &    0.16 \\
\hline
Total                                                 &    0.33    &    1.12    &    0.55    &    0.47    &    0.44 \\
\end{tabular}
\end{ruledtabular}
\end{table*}

A detailed Monte Carlo simulation is used to determine the 
acceptance for each decay mode. 
These acceptances are used 
to correct the background-subtracted numbers of events 
for each partial width ratio.
The acceptance, $A_i$ for decay mode ``$i$'', is defined as
\begin{equation}
    A_i \equiv N^{rec}_i/N^{gen}_i,
\end{equation}
where $N^{gen}_i$ is the number of events generated 
within the nominal $\EK$ and $\ZK$ ranges, and
$N^{rec}_i$ is the number of reconstructed 
events~\footnote{
  For \KLpmz, the acceptance and yield are calculated in 
  five independent \mpp\ bins (20~MeV/$c^2$ per bin),
  and then summed for the measurement of \Gppp/\Gpienu;
  this reduces sensitivity to the decay form factors.
    }.
To account for radiative effects and resolution, which can cause
reconstructed
events to migrate across the $\EK$ and $\ZK$ boundaries, $N_i^{rec}$
is determined from a Monte Carlo generated with broader $E_K$ and
$Z_K$ ranges.

The systematic uncertainties for this analysis, which are summarized in
Table~\ref{tab:syst}, fall into the following categories: acceptance,
background, external branching ratios, and MC statistics.
Uncertainties in acceptance, 
which result from imperfections in the Monte Carlo simulation, 
are by far the most important.

The Monte Carlo simulation includes four main steps, 
each of which introduces systematic uncertainties in the acceptance:  
event generation, 
radiative corrections,
propagation of particles through the detector, 
and detailed simulation of detector response.  
In the following sections, 
we will discuss these MC steps as well as the
associated systematic uncertainties.

  \subsection{Event Generation}
  \label{subsec:mcgen}

\begin{figure}
  \epsfig{file=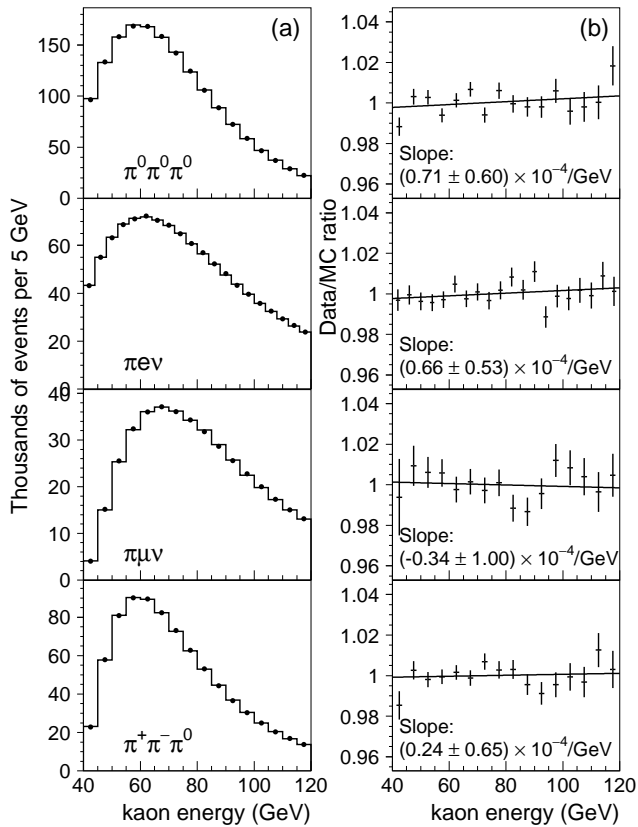, width=\linewidth}
  \caption{
     Comparison of the vacuum beam kaon energy distributions 
     for data (dots) and MC (histogram). 
     For the semileptonic and $\Kthreepi$ modes, 
     the highest $\EK$ solution is plotted.
     The data-to-MC ratios on the right
     are fit to a line, and the $\EK$-slopes are shown.
         }
  \label{fig:ekhi_slopes}
\end{figure}

Event generation starts by selecting a kaon energy
from a spectrum tuned with $10$ million 
\KLpm\ events~\cite{Alavi-Harati:2002ye}. 
Figure~\ref{fig:ekhi_slopes} shows data-to-MC comparisons
of the reconstructed energy 
for the four main \KL\ decay modes.
To limit possible acceptance biases in the determination of 
this spectrum, we compare the high-statistics 
\KLpm and \KLzz\ spectra from the $\epe$ samples.
These spectra agree to better than 1\%
for kaon energies between 40 and 120 GeV (or
0.01\% per GeV).
For each partial width ratio,
the systematic uncertainty is based on
this 0.01\%/GeV $\EK$-slope uncertainty and the 
difference between the average kaon energy of the two decay modes.

The decay kinematics for the 3-body decay modes
depends on form factors. 
For the \KLpmz\  and \KLzzz\ decay modes, 
PDG~\cite{pdg02} values of the form factors with 
2~sigma uncertainties are used. 
For semileptonic form factors we use 
our own measurements~\cite{ktev_kl3ff}. 
The largest uncertainty from form factors is in the \KLpmz\ decay mode.

  \subsection{Radiative Corrections}
   \label{subsec:radcor}

\begin{figure}
  \epsfig{file=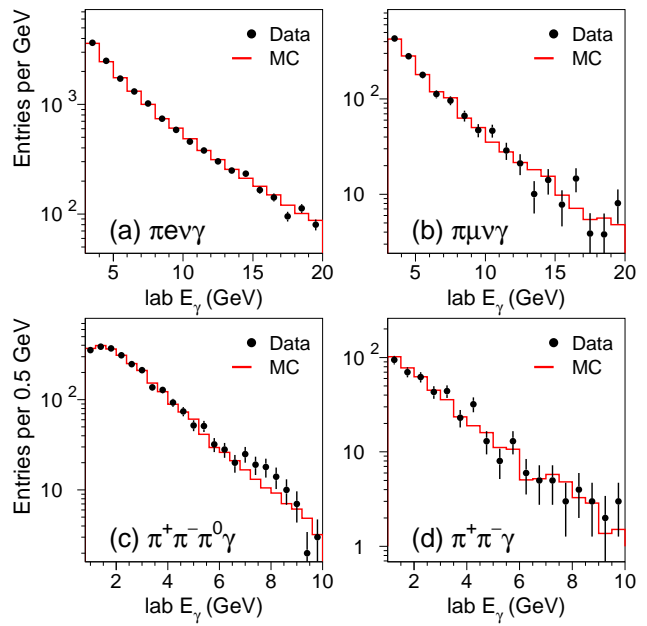, width=\linewidth}
  \caption{
     Comparison of the energy distribution for 
     radiative photon candidates
     for each of the charged decay modes.     
     The data are shown as dots and the MC as a histogram.
         }
  \label{fig:radcor}
\end{figure}

For \Kethree\ and \Kmuthree, inner bremsstrahlung (IB)
contributions are accounted 
for using a new program, {\tt KLOR}, described in~\cite{troy}. 
{\tt PHOTOS}~\cite{photos2} 
is used to generate the IB contribution in \KLpmz\ decays.
The simulation of the $\KLpm$ decay mode includes IB contributions, but
does not include the direct emission 
component, which has a negligible impact on this analysis.

To check the simulation of IB, 
we have performed an analysis 
for each charged decay mode in
which high-energy radiated photons are identified 
in the CsI calorimeter.
Figure~\ref{fig:radcor} shows the data and MC energy distribution for
radiative photon candidates in the laboratory frame;
the shape and normalization agree for all decay modes.

The simulation without IB changes the acceptance by 
2-3\% for the \Kethree\ mode, depending on the electron
energy requirement, 
and less than 0.5\% for the other modes. 
The systematic uncertainty in the partial width ratios is taken
to be 6\% of the acceptance change from IB based on our study
of radiative decays.
The resulting uncertainties vary from 0.14\% to 0.20\%.

  \subsection{Particle propagation}
  \label{subsec:mctrace}

Once a kaon decay is generated, the decay products and their
secondaries are propagated through
the detector.  
This propagation includes both electromagnetic and hadronic
interactions of particles with
detector material, and requires precise modeling of the 
detector geometry and composition.
To model interactions in the detector,
{\sc geant}~\cite{geant} is used to generate process-specific
libraries that are used by our MC.

  \subsubsection{Detector Material}
  \label{subsubsec:material}

\paragraph{Electromagnetic interactions} in detector material 
include  photon
conversions, \brems, multiple
scattering, and $\delta$-ray production. 
For photon conversions and \brems, most particle losses 
in the reconstruction result from interactions upstream of the analysis magnet.
For multiple scattering and $\delta$-ray production, interactions up to
the last drift chamber (DC4) are important.
We estimate 0.73\%\  radiation lengths of material upstream 
of the analysis magnet, 
and a total of 1.18\% radiation lengths of material through DC4.

\begin{figure}
  \epsfig{file=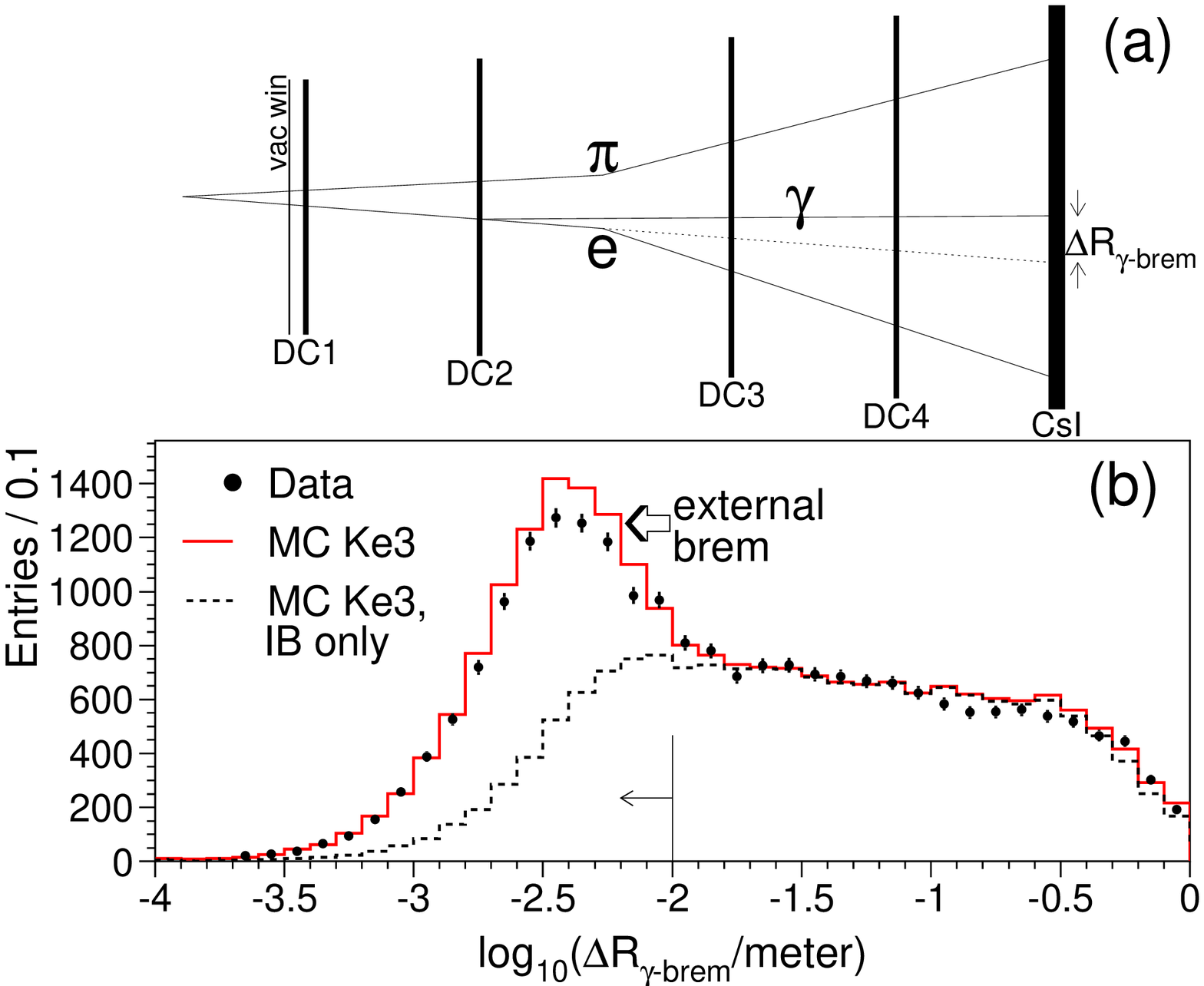, width=\linewidth}
  \caption{
(a) Illustration of electron \brems\ at DC2 for a \KLpienu\ decay.
The dotted line is extrapolated
from the upstream electron trajectory to the CsI.
The track bend between DC2 and DC3 is from the magnet kick.
(b) Measured distribution of \drbrem. 
Data are shown by dots, and MC \Kethree\ by histogram.
The MC contribution from radiative $\KLpienu\gamma$ is shown
by the dashed histogram.
The MC sample is normalized to the total number of
\Kethree\ candidates in data.
         }
  \label{fig:xbrem}
\end{figure}

To check our estimate of detector material upstream of the analysis
magnet, we study $\KLpienu$ decays in which an external \brems\ photon 
is identified in the CsI calorimeter.
In this study,
we take advantage of the magnet kick to separate the electron
from the \brems\ photon.
Figure~\ref{fig:xbrem}(a) illustrates a \Kethree\ decay in which
a photon is produced in DC2.
Figure~\ref{fig:xbrem}(b) shows the distribution of distances (\drbrem )
between the candidate photon cluster and the extrapolation of the
electron trajectory (measured in DC1+DC2) to the CsI calorimeter.
The peak in Fig.~\ref{fig:xbrem}(b) is mainly from external
\brems, while the high-side shoulder is from  
radiative $\KLpienu\gamma$ decays.
To isolate events with external \brems, we require $\drbrem<1$~cm;
the background from radiative $\KLpienu\gamma$ events is 43\%.
The MC sample is normalized to the total number
of \Kethree\ decays in data; 
the fraction of events with a \brems\ photon in this study
is the same in data and MC to within 5\%.

The material downstream of the analysis magnet (including the 0.027 $X_0$
trigger hodoscope) is checked with
\KLzzz\ decays in which
one or more of the photons converts and
gives hits in the hodoscope
\footnote{
        The nominal \KLzzz\ and \KLzz\ analyses allow hits
        in the trigger hodoscope.
    }.
The fraction of events with hodoscope hits is 
measured to be $(13.06 \pm 0.03)\%$
in data and $(12.91 \pm 0.03)\%$ in MC.

The simulation of \dray s and multiple scattering each use
a {\sc geant}-based library. The \dray\ simulation is tuned
to data using DC signals recorded at earlier times than
expected based on the track location in the DC
cell.
To check the \dray\ simulation, we consider the
case in which a \dray\ drifts from one drift chamber cell 
into a neighboring cell
and leaves an extra hit adjacent to the track.
This effect is observed in 35\% of the events in data
(corresponding to about 2\% probability per wire-plane),
and 30\% of the MC events. We therefore assign a 15\%\ uncertainty
to the \dray\ simulation.
The uncertainty in our simulation of 
multiple scattering is estimated 
to be 10\% of its effect on the acceptance,
based on data-MC comparisons of the matching of charged 
particle  trajectories at the decay vertex.

\paragraph{Hadronic interactions} of charged pions can result in
broad  showers that lead to tracking losses and
energy deposits in veto detectors.
Pion interactions in the spectrometer and trigger hodoscope 
are simulated with
{\sc geant}-based libraries.  Interactions in material up through
the last drift
chamber (0.7\%\ pion interaction
lengths) usually result in track loss.  We assign a 
50\% uncertainty on this source of track loss, as described in
Appendix~\ref{app:trackloss}.
The corresponding 0.35\% uncertainty in pion track-loss
affects partial width ratios
with a different number of charged pions in the numerator
and denominator.

Pion interactions in the trigger hodoscope (1.2\% pion 
interaction lengths) downstream of the charged spectrometer
often  prevent the track from matching a cluster in the CsI
calorimeter.
These interactions also can produce
hadronic showers that deposit energy in the \csiveto.
The inefficiency of the track-cluster match requirement for pions
is 0.6\% for data and 0.5\% for MC.
The associated systematic uncertainties in the partial width ratios
are negligible because only one of the two tracks
is required to match a cluster.
The \csiveto\ is only used in the trigger for
$\Gpm/\Gpienu$ (Table~\ref{tb:trigger}); the loss is measured to be 
$(0.4\pm 0.2)$\%\ and is included in the MC.

Most charged pions that do not decay
interact hadronically in the CsI or muon system steel.
The fraction of pions that penetrate the steel and 
produce
a signal in the muon hodoscope 
is measured 
with fully 
     reconstructed \KLpmz\ decays from the low intensity sample;
     the fraction is determined to be 
     $(1.0\pm 0.1)\times 10^{-4}p_{\pi}$,
     where $p_{\pi}$ is the pion momentum in GeV/c.

  \subsubsection{Detector Geometry}
  \label{subsubsec:geom}

\begin{figure}
  \epsfig{file=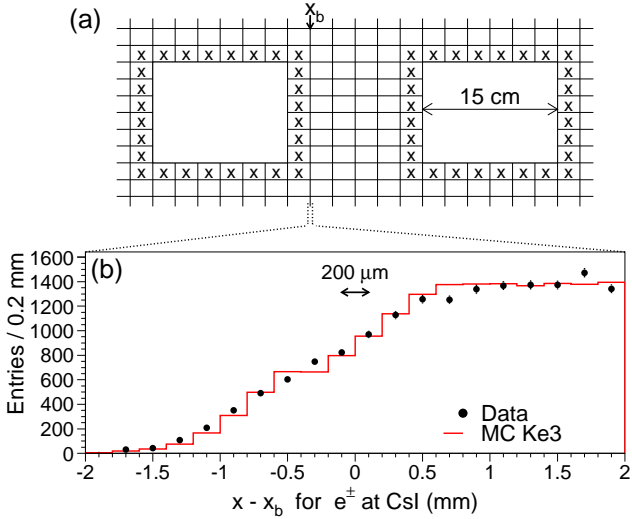, width=\linewidth}
  \caption{
     (a) layout of inner CsI region near beam-holes.
         Photons with reconstructed CsI position in a crystal adjacent 
         to a beam hole (marked with an ``{\sf X}'')
         are rejected in the analysis. $x_b$ is the coordinate
         at one of the crystal boundaries that defines the photon acceptance.
     (b) For \KLpienu\ decays, $x-x_b$ for electrons
         at the CsI determined by extrapolating the trajectory
         measured in the spectrometer.
         The CsI position requirement in (a) is applied
         to this electron sample.  The $200~\mu m$ wide arrow
         indicates the systematic uncertainty for this photon aperture.
         }
  \label{fig:csihole}
\end{figure}

The dimensions of the four drift chambers are known to 
better than 20$\mu$m  based on optical surveys.
The spectrometer is aligned
in situ as explained in~\cite{Alavi-Harati:2002ye}. 
The CsI inner aperture for photons is illustrated in
Fig.~\ref{fig:csihole}(a). 
This aperture is measured by comparing extrapolated
electron tracks measured in the spectrometer with cluster positions 
measured in the calorimeter (Fig.~\ref{fig:csihole}(b));
the uncertainty in this aperture size is $200\mu$m.
The calorimeter dimensions (1.9 m $\times$ 1.9 m)
are known to better than $1$mm.

\begin{figure}
  \epsfig{file=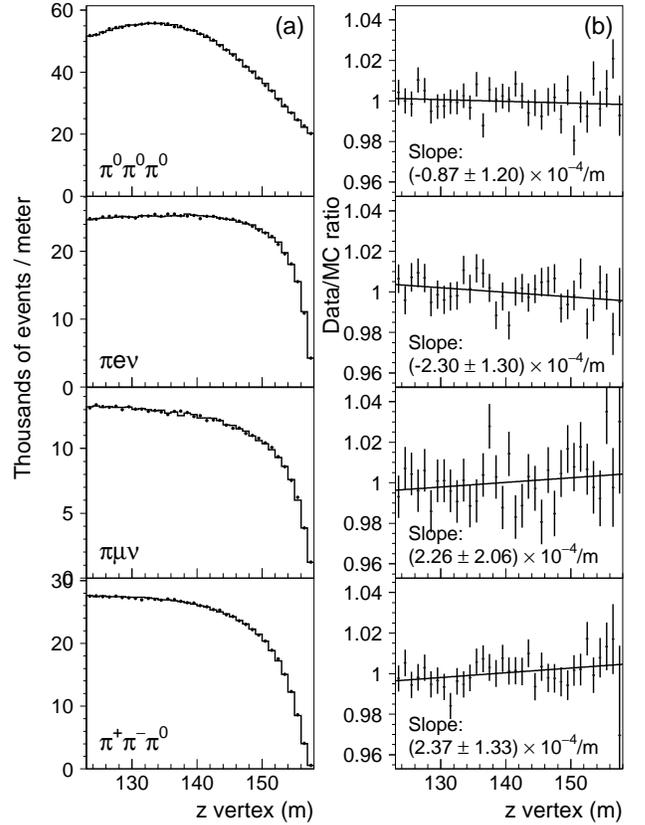, width=\linewidth}
  \caption{
     Comparison of the vacuum beam $z$ distributions for data (dots) 
     and MC (histogram). The data-to-MC ratios on the right
     are fit to a line, and the $z$-slopes are shown.
         }
  \label{fig:zvtx_slopes}
\end{figure}

The agreement of data and MC decay vertex distributions 
(Fig.~\ref{fig:zvtx_slopes}) provides a sensitive
overall check of the detector geometry. 
For each partial width ratio,
the systematic uncertainty is the product of the data/MC
$\ZK$-slope  and the 
difference between the average reconstructed $\ZK$ of the two decay modes.

  \subsection{Detector Response}
  \label{subsec:mcresponse}

The Monte Carlo includes a detailed simulation
of detector response to different particle species,
as well as the effect of accidental activity.

  \subsubsection{Accidentals}
  \label{subsubsec:accid}

Accidental detector activity is measured with a trigger that is proportional
to the instantaneous beam intensity.
We overlay an accidental data event on each generated decay
in the Monte Carlo simulation.
Based on data-MC comparisons of extra detector activity, 
the systematic uncertainty 
is estimated to be $10\%$ of the acceptance change arising from
overlaying accidental data events in the simulation. 
The effects of accidentals result in 0.22\% uncertainty
in the $\Gzzz/\Gpienu$ ratio, and less than 0.04\% in the
other partial width ratios.

  \subsubsection{Trigger}
  \label{subsubsec:trigger}

As described in~\cite{Alavi-Harati:2002ye}, the \ktev\ MC
includes a detailed trigger simulation.
Although the analysis requirements are designed to be
stricter than the trigger, potential anomalies in the trigger
pre-scales and signal processing
could result in losses that are not simulated.
The general strategy to estimate these effects 
is to use \Kethree\ decays from a charged-track trigger 
to study the total CsI energy trigger (Table~\ref{tb:trigger}),
and to use charged decay modes collected in the
total-energy trigger to study the charged-track trigger.

For partial width ratios in which events for 
the numerator and denominator are collected with 
the same trigger, uncertainties in trigger efficiency
largely cancel, and resulting uncertainties are less than 0.1\%.
For $\Gzz/\Gzzz$, the only ratio measured with two separate triggers,
the systematic uncertainty from the trigger efficiency is 0.28\%.

  \subsubsection{Response to Charged Particles}
  \label{subsubsec:mc charged}

Each charged decay mode includes a different combination of 
$e^{\pm}$, $\pi^{\pm}$, and $\mu^{\pm}$.
The simulation of the drift chamber response
includes measured wire inefficiencies,
and several subtle effects that cause non-Gaussian tails
in the position resolution~\cite{Alavi-Harati:2002ye}.
To simulate the response in the CsI calorimeter,
a separate {\sc geant} library is generated 
for $e^{\pm}$, $\pi^{\pm}$, and $\mu^{\pm}$.
Tails in the CsI energy response are measured
in data as described in Appendix~\ref{app:eop_tails}.

Sources of systematic uncertainty are the tracking efficiency,
drift chamber calibration, tails in the $E/p$ distribution,
and analysis cut variations.
The uncertainty in the drift chamber efficiency
is 0.6\% as explained in Appendix~\ref{app:trackloss};
this uncertainty affects only the $\Gzzz/\Gpienu$ ratio.
The DC calibration introduces 
a systematic uncertainty less than 0.1\% on each ratio.
The effect of tails in the CsI energy response
introduces systematic uncertainties well below 0.1\%
on the charged partial width ratios.
Cut variation studies introduce a 0.2\% uncertainty
on all partial width ratios.

  \subsubsection{Response to Photons}
  \label{subsubsec:mc photons}

The CsI calorimeter response to photons is simulated with a
{\sc geant}  library.
The low-side tail in the energy response is assumed
to be the same as for electrons (Appendix~\ref{app:eop_tails}).
The most crucial role of the photon simulation 
is to predict the efficiency of reconstructing
the \KLzzz\ decay mode for the $\Gzzz/\Gpienu$ ratio.
The sources of systematic uncertainty in reconstruction of
multi-$\pi^0$ events are: photon pairing efficiency,
energy scale, and photon reconstruction efficiency.

The pairing efficiency study uses \KLzzz\ decays since
there is no background after identifying six photon clusters 
in the CsI calorimeter.
The sidebands in the 
$3\pz$ invariant mass distribution, as well as events
with \pchi\ values beyond the selection cut,
result from misreconstructed \KLzzz\ events.
These misreconstructed events result almost entirely from
selecting the incorrect photon pairing.
Fig.~\ref{fig:neutmass}(c) illustrates misreconstructed events
in the $3\pz$ invariant mass distribution.
The MC sample in Fig.~\ref{fig:neutmass}(c) is normalized
to data in the 15~MeV wide signal region; the 
data and MC sidebands are in good agreement.
The combined requirements on \pchi\ and $3\pz$ invariant mass
remove $(0.80 \pm 0.01)\%$ of the \KLzzz\ events in data,
and remove $(0.66\pm 0.01)\%$ of the events in MC.
This 0.14\% difference is included as a systematic uncertainty
on the $\Gzzz/\Gpienu$ ratio.

Uncertainties in the calorimeter energy scale and linearity
affect the $\pz$ reconstruction efficiency, primarily
because of the 
photon energy and kaon energy requirements. 
The systematic uncertainty is based on the same set of tests  
as in the $\epe$ analysis; these tests 
result in a 0.33\% uncertainty on the \Gzzz/\Gpienu\ ratio,
and a 0.05\% uncertainty on the \Gzz/\Gzzz\ ratio.

We consider three sources of uncertainty in the photon 
reconstruction efficiency: detector readout, dead material,
and CsI cluster shape requirements.
The calorimeter readout inefficiency is monitored with a laser system,
and is measured to be less than $10^{-6}$.
The amount of dead material between crystals 
is checked with muons, and is included in the MC;
the probability of losing a photon because of this dead
material is less than $10^{-5}$. 
The effect of the photon cluster shape requirement for 
$\gamma$ candidates in the calorimeter is studied by 
removing this cut in the $\Gzzz/\Gpienu$ analysis;
the corresponding change of 0.05\% is taken as a systematic
uncertainty.

  \subsection{Sensitivity to $\pi^0$ Branching Fractions}

For decay modes that use $\pzgg$ decays, 
we  correct for the branching ratio, 
$B(\pzgg) = 0.9880 \pm 0.0003$.
The decay $\pz\to\eeg$ has a negligible effect on all
decay mode acceptances except for \KLpmz.
Although the measurement of \KLpmz\ ignores the \pz,
it is still sensitive to the $\pz \to \eeg$ decay
because the extra tracks can cause the event to be rejected.
This effect is studied using a MC sample of \KLpmz\ decays
in which the \pz\ decays to $\eeg$;
the acceptance for this MC sample is $0.32 \times A_{+-0}$,
where $A_{+-0}$ is the nominal acceptance for \KLpmz\ with
\pzgg.
We  assign a $20\%$ uncertainty to the 
fraction of these events passing the selection, 
resulting in a $0.1\%$ uncertainty
in the \Gppp/\Gpienu\ partial width ratio.

\section{Results}
\label{sec:results}

\subsection{Partial width ratios}

The numbers of events and 
detector acceptances for all decay modes 
are summarized in Table~\ref{tab:ratios};
the resulting partial 
width ratios
are given in the last column of the table.
For each partial width ratio, the first error is statistical and the second
systematic.
The systematic uncertainty is calculated as
the sum in quadrature of the individual sources. 
Note that 
although the partial width ratios use independent data
samples, there are correlations among the systematic errors.
For example,
uncertainties from external bremsstrahlung cause correlated uncertainties 
among the four partial width ratios involving $\Gpienu$.
These correlations are
treated as described in Appendix D of Ref.~\cite{Alavi-Harati:2002ye}.
The correlation coefficients for the five partial width ratios are
given in Table~\ref{tb:corr}.

The systematic precision of each ratio depends on the cancellation 
of systematic uncertainties for the pair of modes. 
Among the partial width ratios, 
the measurement of \Gzzz/\Gpienu\ has the largest
systematic uncertainty, with $\sigma_{syst} = 1.12\%$.
This ratio has the largest uncertainty because there is no 
cancellation of the 0.6\% uncertainty in the tracking efficiency,
as well as the 0.37\% uncertainty in the $3\pz$ reconstruction.
Pairs of modes with  the same number of charged pions
in the final state (\Gpimunu/\Gpienu\ and \Gzz/\Gzzz) 
have the smallest systematic uncertainty, with 
$\sigma_{syst} \sim 0.4\%$.
Pairs of modes in which the number of charged pions is different 
(\Gppp/\Gpienu\ and \Gpm/\Gpienu) have a larger 
systematic uncertainty of $\sigma_{syst} \sim 0.55\%$;
this increased uncertainty is mainly from the 0.35\%
uncertainty in losses from hadronic interactions.

\begin{table*}
\caption{
    \label{tab:ratios} Background-subtracted numbers of events, 
     detector acceptances, and
     resulting ratios of partial decay widths. The $\times x$ next to 
     a number of events reflects a prescale that must be applied to calculate
     the partial width ratio.
     For the partial width ratios, the first error is statistical and the
     second systematic. Note that differences in event selection 
     requirements for the partial width ratios result in a range of 
     acceptances for \KLpienu.
       }
\begin{ruledtabular}
\begin{tabular}{ lccc}
Decay Modes       & ~~~Numbers of Events~~~ & ~~~Acceptance~~~ & Partial Width Ratio \\
\hline 
$\Gpimunu/ \Gpienu$  & $\NdataPMNa / \NdataPMNb $ & \accPMNa / \accPMNb &  $\RPMNvalue \pm \RPMNerrstat \pm \RPMNerrsyst $\\
$\Gzzz/ \Gpienu$  & $\NdataZZZa/ \NdataZZZb$  & \accZZZa / \accZZZb & $\RZZZvalue \pm \RZZZerrstat \pm \RZZZerrsyst $\\ 
$\Gppp/ \Gpienu$  & $\NdataPPPa/ (\NdataPPPb \times 2)$ &  \accPPPa / \accPPPb
 & $\RPPPvalue \pm \RPPPerrstat \pm \RPPPerrsyst $\\ 
$\Gpm/ \Gpienu$  & $\NdataPPa/ (\NdataPPb \times 8)$ &   \accPPa / \accPPb
& $(\RPPvalue \pm \RPPerrstat \pm \RPPerrsyst) \times 10^{-3} $\\ 
$\Gzz/ \Gzzz$  & $\NdataNEUTa/ (\NdataNEUTb \times 5)$ & \accNEUTa / \accNEUTb
& $(\RNEUTvalue \pm \RNEUTerrstat \pm \RNEUTerrsyst ) \times 10^{-3}$\\ 
\end{tabular}\end{ruledtabular}
\end{table*}

  \subsection{Cross Checks of Partial Width Ratios}
  \label{subsec:crosscheck}

We have performed several crosschecks of our measurements.
Some of these checks affect all of the partial width ratios,
while others are relevant to a specific decay mode.

\begin{figure}[hb]
 \epsfig{file=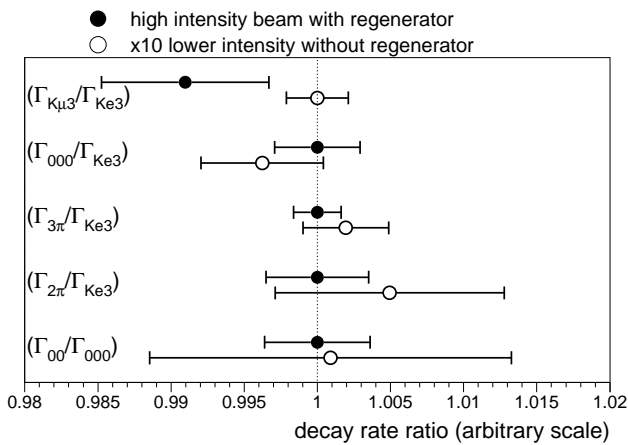, width=\linewidth}
\caption{
    \label{fig:intensity} 
       Partial width ratios measured with high intensity (solid dots)
       and with low intensity (open circles).
       Each ratio is normalized so that the quoted result is one.
       The error bars reflect the statistical uncertainties
       between the two samples.
         }
\end{figure}

As discussed in Sec.~\ref{sec:data}, 
we measure each partial width ratio in 
both high and low intensity data samples.  
The factor of 10 difference in beam intensity between these samples 
results in significantly different tracking and 
photon cluster reconstruction efficiencies.
For example, the tracking inefficiency is nine times smaller in the
low intensity sample (0.38\%\ vs. 3.3\%).
Figure~\ref{fig:intensity} compares the partial width ratios measured in
these two samples. 
The measurements are in good agreement: 
the $\chi^2$ per degree of freedom is $\lovshich/5$.

To check our nominal \KLpimunu\ analysis, 
which does not use the muon system,
we perform an analysis that requires one track to be matched to hits 
in the most downstream
muon hodoscope.
Since the muon track is identified, there are only 2 kaon energy solutions
instead of the 4 solutions in the nominal analysis.
This alternate analysis differs from the nominal analysis by
$(0.08 \pm 0.02_{\mbox{stat}})\%$.
It is worth mentioning that requiring a signal in the 
muon hodoscope does not significantly reduce the background
because \KLpmz\ and \KLpm\ decays can pass the \Kmuthree\
selection only if one of the pions decays
in flight.

The nominal \KLpmz\ analysis, which does not reconstruct the $\pz$,
is checked by performing an analysis in which the $\KLpmz$ decay is fully
reconstructed using the \pzgg\ decay.
Requiring a reconstructed \pzgg\ decay in the CsI calorimeter
reduces the acceptance by a factor of four.
To increase the statistical significance of this crosscheck,
we use an independent sample (with $\times 5$ smaller pre-scale) 
collected in the trigger used to measure $\Gpm/\Gpienu$
(see Table~\ref{tb:trigger});
the two methods agree to $(0.03 \pm 0.28_{\mbox{stat}})\%$.

The stability of the results is also tested by dividing the
data into a variety of subsamples based on criteria such as
vertex position, kaon energy, minimum track separation, 
and minimum photon energy.
The measured partial width ratios are found to be consistent 
within the uncorrelated statistical uncertainty in all of these
studies.

The clean separation of semileptonic and \KLpmz\ decays in 
the \ppzkin\
distribution (Fig.~\ref{fig:rawkin}(c)) allows a 
measurement of $R_{+-0} = \Gppp/(\Gpimunu + \Gpienu + \Gppp)$ that 
does not use any particle identification information. 
The difference between this fit and the nominal analysis is
$(0.35 \pm 0.51)\%$.

Assuming lepton universality, we can make
an independent prediction of the $\Gpimunu/\Gpienu$ ratio:
\begin{equation}
 \Big[\frac{\textstyle \Gpimunu}
 {\textstyle \Gpienu}\Big]_{\textstyle pred} = \frac{\textstyle 1+\delta_K^\mu  }
       {\textstyle 1 + \delta_K^e}
                \cdot
    \frac{\textstyle I_K^{\mu}}
         {\textstyle I_K^{e}}. 
    \label{eq:universal}  
\end{equation}
Here,  $\delta_K^\ell$  represents the mode-dependent long-distance
radiative correction
to the  total decay width, and $I_K^{e}$ and $I_K^{\mu}$ 
are the decay phase space integrals, 
which depend on the form factors. Using
the KTeV measurement of  $\textstyle I_K^{\mu}/\textstyle I_K^{e} =
\ikrbv \pm \ikrbetot$~\cite{ktev_kl3ff} and
$(1+\delta_K^{\mu})/(1+\delta_K^e) = 
\DeltaRat \pm \DeltaRatE$ from KLOR~\cite{troy},
we find that the ratio of the directly measured to predicted values
of $\Gpimunu/\Gpienu$ is
$\LeptUni \pm \LeptUniE$, consistent with 1.

\subsection{ Determination of Branching Fractions, 
             Partial Decay Widths, 
             and \mageta\  }

\begin{table*}[ht]
\caption{
   \label{tb:corr}
   Total uncertainties and correlation coefficients for 
   the partial width ratios.
        }
\begin{ruledtabular}
\begin{tabular}{lccccc}
  & \Gpimunu/ \Gpienu  & \Gzzz/ \Gpienu  
& \Gppp/ \Gpienu & \Gpm/ \Gpienu & \Gzz/ \Gzzz  \\
\hline
Total Error  &  \RPMNerrtot
             &  \RZZZerrtot
             & \RPPPerrtot
             & $\RPPerrtot \times 10^{-3}$
             & $\RNEUTerrtot \times 10^{-3}$       \\ \hline 
%
%
~ & \multicolumn{5}{c}{Correlation coefficients} \\

\Gpimunu/ \Gpienu    &  1.00      &       &       &       &       \\
\Gzzz/ \Gpienu       &  0.14 &  1.00      &       &       &       \\
\Gppp/ \Gpienu       &  0.21 & -0.06 &  1.00      &       &       \\
\Gpm/ \Gpienu        &  0.24 & -0.07 &  0.49 &  1.00      &       \\
\Gzz/ \Gzzz          &  0.09 &  0.30 &  0.04 &  0.07 &  1.00      \\
\end{tabular}
\end{ruledtabular}
\end{table*}

\begin{table}
\caption{
    \label{tab:br}
     $K_L$ branching fractions and partial widths ($\Gamma_i$).
     The partial width measurements
     use the PDG average for the $K_L$ lifetime: 
     $\tau_L= (\TLvalue \pm \TLerror) \times 10^{-8}$~sec~\cite{pdg02}.
     The quoted errors are the sum in quadrature of statistical and
     systematic uncertainties.
       }
\begin{ruledtabular}
\begin{tabular}{ lcc}
Decay Mode       & Branching Fraction & $\Gamma_i$ ($10^7 s^{-1}$) \\
\hline 
\KLpienu & $\BKEvalue \pm \BKEerrtot $ & $\WKEvalue \pm \WKEerrtot $ \\
\KLpimunu & $\BKMvalue \pm \BKMerrtot  $&  $\WKMvalue \pm \WKMerrtot  $ \\
\KLpmz & $\BPMZvalue \pm \BPMZerrtot $ & $\WPMZvalue \pm \WPMZerrtot $  \\
\KLzzz & $\BZZZvalue \pm \BZZZerrtot  $ & $\WZZZvalue \pm \WZZZerrtot  $ \\
\KLpm & $(\BPMvalue \pm \BPMerrtot)\times 10^{-3}  $ 
       &  $(\WPMvalue \pm \WPMerrtot)\times 10^{-3}  $  \\
\KLzz & $(\BZZvalue \pm \BZZerrtot)\times 10^{-3}  $ 
& $(\WZZvalue \pm \WZZerrtot)\times 10^{-3} $ \\
\end{tabular}
\end{ruledtabular}
\end{table}

\begin{table*}[htb]
\caption{
   \label{tb:corr2}
   Total uncertainties and correlation coefficients for 
   the $K_L$ branching  ratios.
        }
\begin{ruledtabular}
\begin{tabular}{lcccccc}
             & B(\KLpienu) & B(\KLpimunu) & B(\KLzzz)  
& B(\KLpmz) & B(\KLpm) & B(\KLzz)  \\
\hline
Total Error  &  \BKEerrtot
             &  \BKMerrtot
             &  \BZZZerrtot
             & \BPMZerrtot
             & $\BPMerrtot \times 10^{-3}$
             &  $\BZZerrtot \times 10^{-3}$             \\ \hline 
%
%
~ & \multicolumn{6}{c}{Correlation coefficients}\\
$B(\KLpienu)$        &  1.00      &       &       &       &       &       \\
$B(\KLpimunu)$       &  0.15 &  1.00      &       &       &       &       \\
$B(\KLzzz)$          & -0.77 & -0.62 &  1.00      &       &       &       \\
$B(\KLpmz)$          &  0.18 &  0.08 & -0.54 &  1.00      &       &       \\
$B(\KLpm)$           &  0.28 &  0.22 & -0.48 &  0.49 &  1.00      &       \\
$B(\KLzz)$           & -0.72 & -0.54 &  0.89 & -0.46 & -0.39 &  1.00      \\

\end{tabular}
\end{ruledtabular}
\end{table*}

Imposing the constraint that the sum of the six largest branching
fractions is 0.9993, we determine
the branching fractions shown in Table~\ref{tab:br}. 
Correlations among the partial decay width 
measurements (Table~\ref{tb:corr}) are taken into account in calculating
uncertainties in the branching fractions. 
The correlation coefficients for
the six branching fractions are given in 
Table~\ref{tb:corr2}.
Using the PDG average for the neutral kaon 
lifetime~\footnote
{For $\tau_L$, we use the PDG average~\cite{pdg02} rather than the PDG fit value to
avoid correlations with $K_L$ branching fractions.},
$\tau_L = (\TLvalue \pm \TLerror) \times 10^{-8}$~sec, 
these branching fractions correspond to the partial widths
quoted in the same table.

The $\KLpp$ measurements, combined with the kaon lifetimes, 
also provide
a precise measurement of $|\etapm|^2 \equiv \Gamma(\KL \to \pi^+\pi^-)/
\Gamma(\KS \to \pi^+\pi^-)$:
\begin{equation}
   \mageta^2 = 
    \frac{\tau_S}{\tau_L} 
    \frac{ \BLpm + \BLzz[1+6Re(\epe)] } { 1 - \BSklthree}, 
     \label{eq:etapm}
\end{equation}
where 
$\BLpm$ and $\BLzz$ are the
$\KLpp$ branching fractions
quoted in Table~\ref{tab:br},
$\tau_S = (0.8963\pm 0.0005)\times10^{-10}$~sec
\footnote{We use average $\tau_S$ from \ktev\ and NA48.},
and $Re(\epe)= (16.7 \pm 2.3)\times 
10^{-4}$ \cite{pdg02,na4802,Alavi-Harati:2002ye}.
We calculate the $K_S\to \pi\ell\nu$ branching fraction 
$\BSklthree= 0.118\%$ assuming that
$\Gamma(K_S\to\pi\ell\nu) = \Gamma(K_L\to\pi\ell\nu)$.
The resulting value of $\mageta$ is
\begin{equation}
   \mageta     =  (\etapmv \pm \etapme)\times 10^{-3}.
   \label{eq:etapm_value}
\end{equation}
The uncertainty in $\tau_L$ contributes $\etapmext\times 10^{-3}$ 
to the uncertainty in
\mageta\, while our $\KL \to \pi\pi$ branching fraction measurements 
contribute an uncertainty of $\etapmint\times 10^{-3}$;
$\tau_S$ and $\reepoe$ contribute negligibly to the error in $|\etapm|$.



  \section{Comparison with Previous Partial Width and
  $|\eta_{+-}|$ Measurements}
  \label{sec:previous}

\begin{figure}[ht]
 \epsfig{file=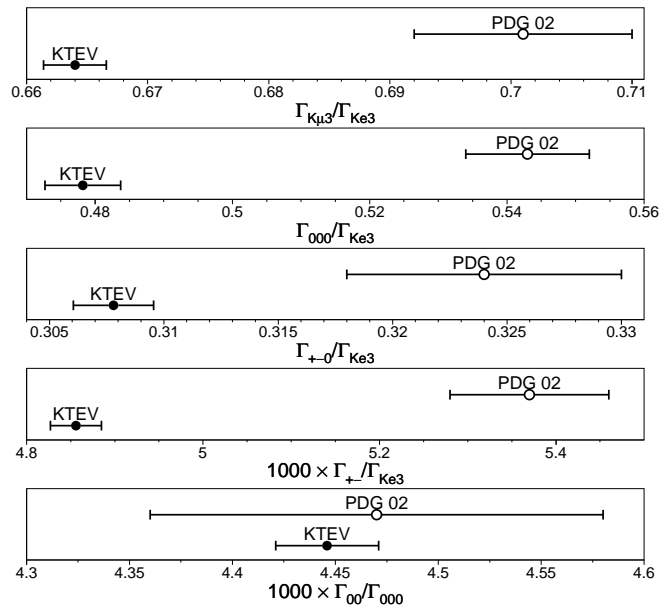, width=\linewidth}
\caption{
    \label{fi:pdg1} 
       Partial width ratios measured by KTeV (dots) and from PDG
       fit (open circles).}
\end{figure}
\begin{figure}[ht]
 \epsfig{file=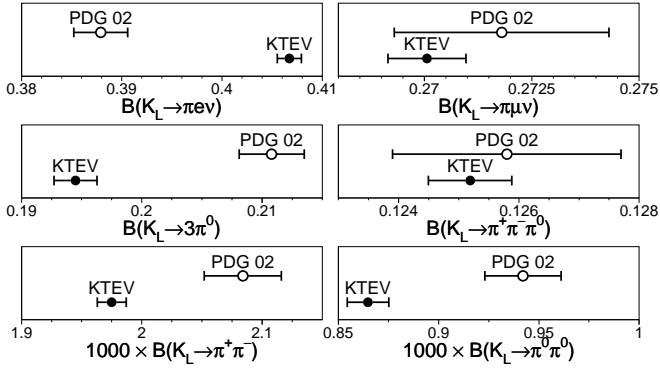, width=\linewidth}
\caption{
    \label{fi:pdg2} 
       $K_L$ branching fractions measured  by KTeV (dots) and from PDG
       fit (open circles).
       }
\end{figure}

The new KTeV measurements of the partial width ratios and $\KL$
branching fractions are on average a factor of two more precise 
than the current world average values, but are not in good agreement
with these averages.
Figure~\ref{fi:pdg1}
shows a comparison of the KTeV and PDG values for the five partial 
width ratios.
Of the five partial width ratios,
only the $\Gzz/\Gzzz$ measurement is in good agreement;
note that
$\Gzz/\Gzzz$ is the only ratio that
does not include the \KLpienu\ decay mode.
Our  measurements  of  \Gpimunu/\Gpienu\ and \Gppp/\Gpienu\ disagree
with the PDG by 
$5\%$, and our measurements of \Gzzz/\Gpienu\ and \Gpm/\Gpienu\ disagree
with the PDG by $10\%$. 
Figure~\ref{fi:pdg2} shows the corresponding comparison of KTeV
and PDG branching fractions.
The discrepancies between KTeV and the PDG can be 
reduced significantly by applying a 7\%\ relative shift to
either the KTeV or PDG values for
B(\KLpienu).

Another measurement, which has not yet been included 
in the PDG summary, is the KLOE measurement of
$R_S^\pi = \Gamma(K_S\to\ppc)/\Gamma(K_S\to\zz)
= 2.236\pm 0.015$~\cite{kloe:kspmzz}.
The \ktev\ measurements of $B(\KLpm)$ and $B(\KLzz)$,
along with the world average value of $\reepoe$, give 
$R_S^\pi = \kspmzz \pm \kspmzze$
in good agreement with, but less precise than, the KLOE result.

\begin{figure}[ht]
 \epsfig{file=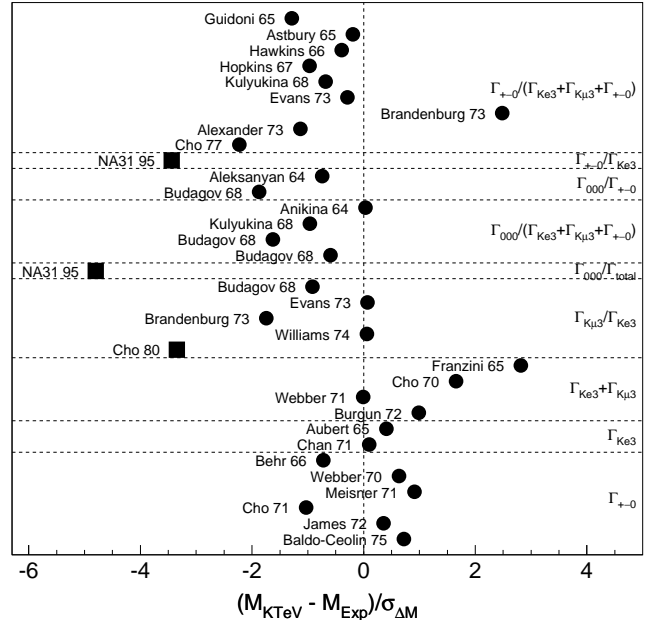, width=\linewidth}
\caption{
    \label{fi:pdg3}
Residual distribution between 34 individual measurements
(${\rm M_{Exp}}$)
used in the PDG fit and the KTeV determinations of these
quantities (${\rm M_{KTeV}}$). $\sigma_{\rm \Delta M}$ is the uncertainty
on ${\rm M_{KTeV}-M_{Exp}}$.
Measurements disagreeing with KTeV
by more than $3 \sigma$ are shown as solid squares.
The measurements are 
{\bf Guidoni 65}~\cite{Guidoni65},
{\bf Astbury 65}~\cite{Astbury65},
{\bf Hawkins 66}~\cite{Hawkins66},
{\bf Hopkins 67}~\cite{Hopkins67},
{\bf Kulyukina 68}~\cite{Kulyukina68},
{\bf Evans 73}~\cite{Evans73},
{\bf Brandenburg 73}~\cite{Brandenburg73},
{\bf Alexander 73}~\cite{Alexander73},
{\bf Cho 77}~\cite{Cho77},
{\bf NA31 95}~\cite{NA3195},
{\bf Aleksanyan 64}~\cite{Aleksanyan64},
{\bf Budagov 68}~\cite{Budagov68},
{\bf Anikina 64}~\cite{Anikina64},
{\bf Williams 74}~\cite{Williams74},
{\bf Cho 80}~\cite{Cho80},
{\bf Franzini 65}~\cite{Franzini65},
{\bf Cho 70}~\cite{Cho70},
{\bf Webber 71}~\cite{Webber71},
{\bf Burgun 72}~\cite{Burgun72},
{\bf Aubert 65}~\cite{Aubert65},
{\bf Chan 71}~\cite{Chan71},
{\bf Behr 66}~\cite{Behr66},
{\bf Webber 70}~\cite{Webber70},
{\bf Meisner 71}~\cite{Meisner71},
{\bf Cho 71}~\cite{Cho71},
{\bf James 72}~\cite{James72},
{\bf Baldo-Ceolin 75}~\cite{Baldo75},
       }
\end{figure}

To understand the discrepancy between KTeV and the PDG averages, we have 
considered the 49 measurements and fit results used in the PDG 
averages~\cite{pdg02}. 34 of these measurements involve decay modes
with branching fractions greater than $1\%$.
Figure~\ref{fi:pdg3} shows the 
distribution of residuals (normalized by uncertainty)
between these measurements and values obtained 
using the new KTeV measurements.    
The $\chi^2$ per degree of freedom is $82.9/34$ showing
a clear inconsistency. 
Approximately 40 units of the $\chi^2$ come from the 
three measurements with greater than 3$\sigma$ disagreement with KTeV:
the measurement
of $\Gpimunu/\Gpienu$ reported by Cho~80~\cite{Cho80}, and
the measurements of  \Gzzz\ and  \Gppp/\Gpienu\ from 
NA31~\cite{NA3195}. It is interesting to note that
in the same paper, NA31 reports $\Gzzz/\Gppp = 1.611\pm 0.037$, 
not involving the $\KLpienu$ decay mode, which is consistent
with KTeV's measurement of $\Gzzz/\Gppp = 1.567 \pm 0.020$; this NA31 
measurement is not used in the PDG branching ratio fit because it is
not independent of the other NA31 measurements included in the fit.

\begin{figure}[ht]
 \epsfig{file=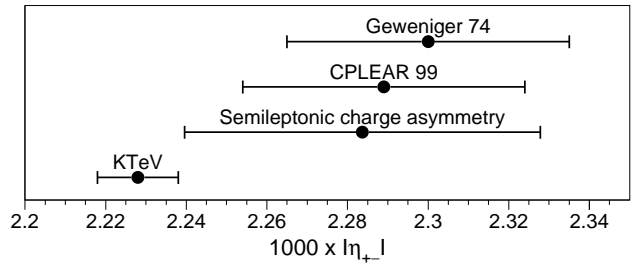, width=\linewidth}
\caption{
    \label{fig:etapm} 
     Comparison of KTeV's \mageta\ measurement with
     previous measurements based on $\K_L$-$\K_S$ 
     interference~\cite{aspk74b,cplear99},
     and the semileptonic charge asymmetry~\cite{ktev_ke3asym,
     pdg02}.
     Note that the CPLEAR value of \mageta\
       has been adjusted to $\tau_S = 0.8963\times10^{-10}$~sec
       using their quoted dependence on the $K_S$ lifetime.
              }
\end{figure}

Figure~\ref{fig:etapm} compares our new determination of \mageta\
with the two measurements based on $K_L$-$K_S$ interference
\footnote{We ignore the PDG fit for \mageta\ because it uses
          the $\KLpp$ branching fractions. The disagreement of
          the KTeV and PDG values for these branching fractions has
          already been discussed.}.
The average of these two previous measurements gives
$\mageta = (2.295 \pm 0.025) \times 10^{-3}$, 
which disagrees with the \ktev\ evaluation by $2.7\sigma$. 
Figure~\ref{fig:etapm} also shows \mageta\ determined from the 
charge asymmetry assuming CPT invariance~\footnote{\mageta\
is related to $\delta_{\ell}$, the $\KLpilnu$ charge asymmetry:
$\mageta = (0.5\,\delta_\ell/\cos\phi_{+-}) + |\epsilon^{\prime}|$, where
$\delta_\ell = (3307 \pm 64) \times 10^{-6}$ is the average of the recent KTeV
measurement~\cite{ktev_ke3asym} and
previous measurements~\cite{pdg02}, $\phi_{+-} = (43.51 \pm 006)^\circ$ 
is the superweak phase~\cite{pdg02}, and $|\epsilon^{\prime}| =
3.8 \times 10^{-6}$~\cite{pdg02}.}; 
the value is consistent with all other
measurements.

  \section{Conclusions}
  \label{sec:conclude}


In summary, we have measured the branching fractions for the six 
$K_L$ decay modes with
branching fractions greater than $0.05\%$. 
The new measurements are about a factor of two more precise than current
world averages, but are not in good agreement with these averages.
Compared to the PDG fit~\cite{pdg02},
the KTeV measurement of $B(\KLpienu)$
is higher by 5\%, $B(\KLzzz)$ is lower by 8\%,
$B(\KLpm)$ is lower by 5\%,  and $B(\KLzz)$ is lower by 8\%.
Our measurements of $B(\KLpimunu)$ and $B(\KLpmz)$ are consistent
with the PDG fit.

The new $K_L$ branching fractions will require the adjustment of
several rare $\KL$ branching fractions for which the main $\KL$ decay modes
are used as normalization.
The $\KL$ branching fraction measurements also may be used
to determine several parameters, including $|\etapm| = 
(\etapmv \pm \etapme)\times 10^{-3}$,
reported
in this paper, and $|V_{us}|$, described in~\cite{ktev_vus}. 

\section{Acknowledgments}

We gratefully acknowledge the support and effort of the Fermilab
staff and the technical staffs of the participating institutions for
their vital contributions.  This work was supported in part by the U.S. 
Department of Energy, The National Science Foundation, and The Ministry of
Education and Science of Japan.

  \appendix

  \section{Determination of Tracking Losses}
  \label{app:trackloss}

The determination of the absolute track reconstruction efficiency is 
most important for the measurement of \Gzzz/\Gpienu\, which compares
a decay mode with two charged 
tracks to a mode without charged particles
in the final state. 
The causes of tracking loss fall into four categories:
(i) missing or corrupted hits induced by accidental activity,
$\delta$-rays, and non-Gaussian tails in the
chamber response; (ii) many spurious  hits, 
mainly from accidental activity, which confuse pattern recognition;
(iii) failure of track segments to satisfy  matching criteria
at the analysis magnet or decay vertex because of
large angle scattering and $\pi\to\mu\nu$ decays;
(iv) hadronic interactions in the spectrometer.
All of these effects
are included in the Monte Carlo simulation,
as described in~\cite{Alavi-Harati:2002ye}. 
Pion decay is the main source of track loss,
and is well described in the MC. 
This appendix describes an inclusive study
of all other sources of track loss.
The first three sources of track loss are related to the
drift chamber performance, and are described in
Sec.~\ref{subsec:track ineff}. 
The effect of hadronic interactions
is described in Sec.~\ref{subsec:hadspec}.

The measurement of the tracking loss is based on partially 
reconstructed \KLpmz\ decays that are identified 
by the $\pzgg$ decay along with one or two 
additional hadronic clusters in the CsI calorimeter 
(i.e, clusters with transverse profile inconsistent with a photon). 
Ideally, \KLpmz\ decays could be cleanly identified using only
the four clusters in the CsI calorimeter,
leading to a direct measurement of the two-track inefficiency;
unfortunately, this sample has significant background.
To obtain a \KLpmz\ sample with low enough background
for this study,
we also require either a completely reconstructed pion track 
or some reconstructed track segments.
With this partial reconstruction,
the missing track information can be predicted from 
kinematic constraints, and then compared with the
track information found by the track reconstruction.
The data used for this study were collected in a trigger
that requires energy in the CsI calorimeter and rejects
events with activity in the muon system.

The two-track inefficiency is measured in two steps.
First, we measure the single track inefficiency after
one track is reconstructed ($\eta_1$).
Next, we measure the probability of finding exactly zero tracks
($\eta_0$). The corresponding two-track loss, $\eta_2$,
is then given by $2\eta_1 + \eta_0$.

  \subsection{Tracking Inefficiency}
  \label{subsec:track ineff}

\begin{figure}[hb]
 \epsfig{file=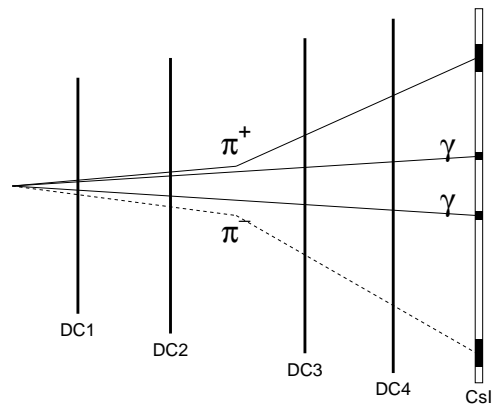, width=\linewidth}
\caption{
    \label{fig:pmz_display} 
       \KLpmz\ topology used to measure single track inefficiency
        ($Z$-$X$ projection).
       Solid lines represent particles reconstructed in the
       detector. The dashed line indicates the pion trajectory
       calculated by the kinematics of the reconstructed particles 
       with the assumption of a \KLpmz\ decay.
       }
\end{figure}

The study of the single track efficiency uses events with
a reconstructed $\pzgg$ along with two hadronic clusters in the 
CsI calorimeter.
Requiring exactly two hadronic clusters suppresses
\KLpmz\ decays with hadronic interactions in the spectrometer
because such events
tend to create additional hadronic clusters.
Events with $\pi \to \mu \nu$ decays are also suppressed because
these events leave only one hadron cluster.
One of the two hadron clusters must be matched to a fully 
reconstructed track in the spectrometer,
leaving two possible kinematic solutions for the missing track.
The fully reconstructed track, 
along with the measured position of the other hadron cluster 
(associated with the missing track),
provides sufficient information to select the correct
kinematic solution for the  missing pion trajectory.
Figure~\ref{fig:pmz_display} illustrates this selection.
For the high intensity sample, we find the single track inefficiency
($\eta_1$) to be $(1.47 \pm 0.02)\%$ in data and $(1.32\pm 0.02)\%$ in MC;
in the low intensity sample, $\eta_1 = (0.18 \pm 0.02)\%$ in data
and $(0.13 \pm 0.02)\%$ in MC.

\begin{figure}[hb]
 \epsfig{file=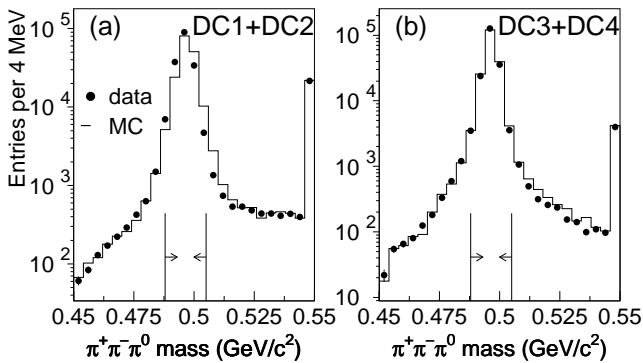, width=\linewidth}
\caption{ 
       For $\eta_0$ tracking inefficiency study,
       $\pmz$ invariant mass distributions 
       after all analysis requirements except $\pmz$-mass.
       The data are shown as dots and the MC as a histogram.
       Distributions are based on identifying track segments in 
       (a) DC1+DC2 and (b) DC3+DC4. 
       Events with invariant mass above
       $0.55$~GeV/c$^2$ are shown in the last bin.
       The horizontal arrows indicate the region selected by the
        \pmz\ mass requirement. The rightmost bin shows 
       all events with mass greater than 0.55 GeV/$c^2$.
     }
    \label{fig:trmass} 
\end{figure}

In the above study, the reconstructed track requirement excludes 
the case in which correlated hit losses within a single drift chamber
result in no reconstructed tracks.
To account for correlated losses within a drift chamber,
we perform a separate analysis to measure the fraction of events
with zero reconstructed tracks ($\eta_0$).
We start with the same calorimeter selection of 
two photon clusters for $\pzgg$ and two hadron clusters.
Next, we require two reconstructed track segments, 
either both upstream of the analysis magnet in DC1 and DC2,
or both downstream of the analysis magnet in DC3 and DC4. 
Figure~\ref{fig:trmass} shows the $\pmz$ invariant mass distributions
for these two samples. 
The $\pmz$-mass resolution for the DC1-DC2 (DC3-DC4) 
selection is about 3~MeV/$c^2$ (2~MeV/c$^2$) compared to $1$~MeV/c$^2$ 
for the standard reconstruction. 
The tails are well described by the $\KLpmz$ MC.
For the high intensity sample, we find 
$\eta_0 = (0.35 \pm 0.02)\%$ in data and $(0.06 \pm 0.02)\%$  in MC.
For the low intensity sample,
$\eta_0 = (0.03\pm 0.01)\%$ in data and zero in MC.

\begin{figure}[hb]
 \epsfig{file=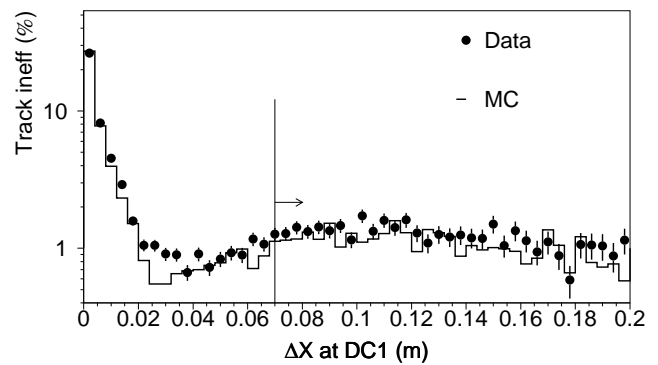, width=\linewidth}
\caption{
   Single track reconstruction inefficiency ($\eta_1$)
   as a function of the $X$-distance ($\Delta X$)
   between the reconstructed track and 
   the calculated track at DC1.
   The arrow indicates the selection requirement.
     }
    \label{fig:separ} 
\end{figure}

In the analyses for both $\eta_1$ and $\eta_0$,
we ensure that the special selection does
not introduce biases in the tracking efficiency,
and that the efficiency corrections 
are applicable for the nominal selection of charged decay modes.
One of the most important aspects in the tracking inefficiency study
is to require separation between the reconstructed track 
(or track segment) and the kinematically predicted track (or track segment). 
This requirement
is necessary because nearby tracks
have a large tracking inefficiency, which is irrelevant 
for the nominal analysis that uses the track separation cut.
Figure~\ref{fig:separ} shows $\eta_1$ as a function of 
$X$-separation ($\Delta X$) between the two tracks at DC1.
At small track separation, the inefficiency is larger than 20\%.
A requirement of $\Delta X > 7$~cm 
safely removes tracks with small separation.

The total two-track inefficiency ($\eta_2 = 2\eta_1 + \eta_0$)
depends strongly on beam intensity.
For the high intensity period,
$\eta_2 = (3.28\pm 0.04)\%$ in data;
it is not completely reproduced
by the simulation for which $\eta_2 = (2.70\pm 0.05)\%$. 
For the low intensity period, the inefficiency in data is much smaller,
$(0.38\pm 0.06)\%$; 
the MC inefficiency is $(0.26\pm0.05)\%$, which agrees with the data.
Based on these measurements,
we apply a $-0.6\%$ correction to the acceptance for the high intensity
period and a $-0.12$\% correction for the low intensity period.
The uncertainty on this correction is taken as 100\% of the effect:
0.6\% and 0.12\% for the high and low intensity periods, respectively.

  \subsection{Hadronic Interactions in the Spectrometer}
  \label{subsec:hadspec}

From our Monte Carlo simulation, the track loss 
from hadronic interactions in the spectrometer 
is 0.7\% per pion track.
Since these interactions also cause the CsI cluster 
associated with the track to be lost,
we cannot measure the track loss using
the method described in Sec.~\ref{subsec:track ineff}.
To identify hadronic interactions clearly without the
second pion cluster,
we tag pions with an interaction between DC3 and DC4.
We use partially reconstructed $\KLpmz$ decays using
$\pzgg$, one complete pion track, and hits from the second 
pion track in the first three drift chambers.
Note that only one hadronic cluster, 
associated with the complete track, is required.

This study uses low intensity data for which the 
tracking inefficiency has a small effect. 
Events are vetoed if there are extra hits in any of the 
first three drift chambers;
this further suppresses accidental activity and simplifies
the particle trajectory determination.  
Hadronic interactions are tagged
by requiring no hit in the last drift chamber
within $1$~cm of the second track extrapolation; 
according to MC, 90\% of these events are 
due to a hadronic interaction.
In data, $(0.14\pm 0.02)\%$ of events
are tagged as having hadronic interactions;
the corresponding fraction for MC is
$(0.096\pm 0.005)\%$.
The difference between these two fractions
leads to a 50\% uncertainty on the track loss from
hadronic interactions.


  \section{Tails in the CsI Energy Response}
  \label{app:eop_tails}


To measure the partial width ratios,
it is important to understand tails in the CsI energy response
for the four different types of particles detected in
the analysis: electrons, photons, pions, and muons.
The energy response tails affect the absolute detection efficiency,
as well as particle misidentification that leads to background.
In this appendix, we describe these tails, how they are 
determined from data, and how they are simulated.

KTeV's {\sc geant}-based MC
does not include the effects of photo-nuclear
interactions ($\gamma N$).
If a $\gamma N$ reaction occurs during the electromagnetic shower
development, a neutron or charged pion can escape the calorimeter,
resulting in an energy deposit that is too low.
Imperfections in the treatment of dead material (wrapping) 
between crystals also can lead to energy tails that are
not properly modeled.
These effects are empirically modeled and included in the MC
by fitting a function
to the low-side tail in the electron $E/p$ distribution
(Fig.~\ref{fig:pid}(a)). 
The key point in this procedure is to select a sample of 
\KLpienu\ decays 
with low enough background to avoid pion contamination 
for electron $E/p$ values as low as 0.6.
To achieve such electron purity, the pion is required to
satisfy $E/p < 0.3$ to avoid swapping the electron and pion
assignments, and $\mpp > 0.370~\umass$ to reject background from
\KLpmz\ decays.  \KLpimunu\ decays are vetoed by the
muon system. 
With the nominal electron selection requirement of  
$E/p > 0.92$, the inefficiency is measured to be 0.2\%. 
This non-Gaussian tail also affects photons in
the neutral decay modes; the presence of this tail
changes the \KLzzz\ acceptance by 0.7\%.

\begin{figure}[hb]
  \centering
  \epsfig{file=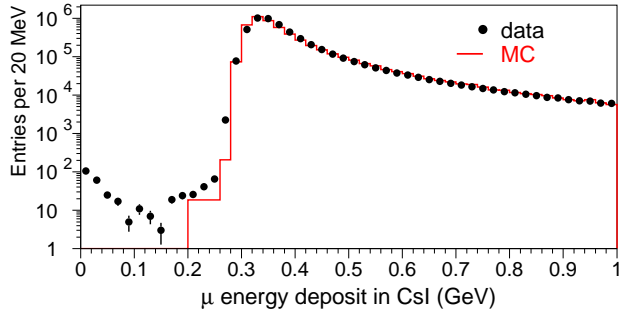,width=\linewidth}
  \caption{ 
   Energy deposit in the CsI for muons with track projection
   at the CsI at least 3~mm from
   any crystal boundary.
     The data are shown as dots and the MC as a histogram.
         }
  \label{fig:emuclus}
\end{figure}

The electron $E/p$ tail in Fig.~\ref{fig:pid}(a)
shows losses up to 40\% of the incident energy
(i.e., down to $E/p \sim 0.6$). To check for
anomalous energy losses of more than 40\%,
we use a sample of muons collected in dedicated runs as 
explained in~\cite{Alavi-Harati:2002ye}. 
We select muons that are at least
3~mm away from any crystal boundary to avoid the known
effects of the small amount of dead material (wrapping)
between crystals.
Figure~\ref{fig:emuclus} shows the energy deposit distribution
for $5.5$ million muons hitting the CsI calorimeter 
away from crystal boundaries.
The fraction of events with an anomalously 
low energy deposit (below 200 MeV) is $5\times 10^{-5}$
in data; 
this effect is negligible in the branching 
fraction analysis.

For charged pions in the CsI calorimeter, 
the $E/p$ distributions for data and {\sc geant}-based MC
agree reasonable well (Fig.~\ref{fig:pid}(b)).
The pion $E/p$ cut inefficiency is measured 
in both data and MC with
\KLpmz\ decays in the low intensity sample.
Note that with the full reconstruction of \KLpmz,
the pion $E/p$ cut is not needed to achieve
a negligible background. 
The inefficiency difference between data and MC 
is measured as a function of the $E/p$ requirement and 
the proximity to the beam holes.
The data-MC inefficiency difference varies between
0.1\% and 0.3\%, and is used to correct the MC samples.

In the \KLpimunu\ analysis, at least one track is required
to deposit less than 2~GeV in the CsI calorimeter (i.e., less
than 5 times the energy deposit of a minimum ionizing particle). 
In a separate \Kmuthree\ analysis that identifies the muon
with the muon hodoscope, the inefficiency of this
cluster energy requirement is measured to be 0.3\% in
both data and MC.
No correction is used to simulate the
muon energy response in the CsI calorimeter.


\end{document}